\documentclass[10pt]{iopart}
\expandafter\let\csname equation*\endcsname=\relax 
\expandafter\let\csname endequation*\endcsname=\relax 

\makeatletter
\def\ps@pprintTitle{%
 \let\@oddhead\@empty
 \let\@evenhead\@empty
 \def\@oddfoot{}%
 \let\@evenfoot\@oddfoot}
\makeatother
\renewcommand{\thetable}{\Roman{table}}
\usepackage{csquotes}
\usepackage{graphicx,color}
\usepackage{float}
\usepackage[caption=false]{subfig}
\usepackage{amsmath}
\usepackage{amssymb}
\usepackage{multirow}
\usepackage{dsfont}
\usepackage{adjustbox}
\usepackage{pdflscape}
\usepackage{wasysym}
\usepackage[percent]{overpic}
\usepackage[dvipsnames]{xcolor}
\usepackage[normalem]{ulem}
\usepackage{makecell} 
\usepackage{array}         
\usepackage[table]{xcolor}

\begin{document}

\title{Thermal resilience of the ITER tungsten first wall to runaway electron impact}
\author{S. Ratynskaia$^{a}$, K. Paschalidis$^{a}$, T. Rizzi$^{a}$, P. Tolias$^{a}$, \\ R. A. Pitts$^{b}$, F. J. Artola$^{b}$, H. Bergström$^c$, V. K. Bandaru$^d$, M. Hoelzl$^c$, S. 
Nicolici$^{b}$}
\address{
$^a$KTH Royal Institute of Technology, Stockholm, SE-100 44, Sweden\\
$^b$ITER Organization, Route de Vinon-sur-Verdon, CS 90 046, 13067 St. Paul Lez Durance Cedex, France\\
$^c$Max Planck Institute for Plasma Physics, Boltzmannstr. 2, 85748 Garching b. M, Germany\\
$^d$Indian Institute of Technology Guwahati, Assam, India}

\begin{abstract}
\noindent The fast volumetric deposition of multi-MeV high current runaway electron (RE) beams constitutes the most critical issue for the ITER tungsten (W) first wall (FW) longevity. Such relativistic electron beams could generate extreme volumetric power densities inside the FW armour which lead to significant vaporization, deep melting and even material explosions, as well as to elevated temperatures at the bond interface with the cooling substrate that could cause rupture and water leaks. Here the thermal response of the ITER FW is modeled with a three-stage, one-way coupled workflow focusing on assessments of the extent of the wall damage and the increase of the bond interface temperature for varying W thickness. Increased W thickness is found to be essential for wall protection against intense RE dissipation events in terms of both W tile damage and cooling system integrity.
\end{abstract}
\maketitle
\ioptwocol

\section{Introduction}\label{sec:introduction}

In ITER, runaway electron (RE) incidence on plasma-facing components (PFCs) poses a major threat to their integrity\,\cite{Pitts_2025, Krieger_2025}. The energetic multi-MeV ITER REs can carry high, up to 10 MA, currents\,\cite{Pitts_2025, Breizman2019, Martin-Solis_2015} and their energy can be deposited in a rather limited volume over relatively short timescales, which can lead to extremely high power densities. This induces a complex thermo-mechanical response, characterized by extreme temperatures, intense vaporization, solid fragmentation or melt splashing and even fast material expulsion\,\cite{Pitts_2025, Ratynskaia_2025b}. In addition to the PFC damage, loss-of-coolant accidents are a major potential concern due to the possibility of overheating the bond interface with the cooling substrate\,\cite{Pitts_2025}. 

It has long been realized that RE dissipation may be a serious concern for the PFC integrity\,\cite{Federici_2001, Bartels1994,Cardella2000,Maddaluno2003,Sizyuk2009,Forster2011,Bazylev2013}, yet our physics understanding of the thermomechanical PFC response to RE incidence has remained rather limited, mainly due to the absence of controlled experiments. Rapid progress has been recently achieved driven by coordinated efforts within the International Tokamak Physics Activity (ITPA) and the EUROfusion consortium. In 2022, the first deliberate well-diagnosed experiments with instrumented ATJ graphite samples were performed in the DIII-D tokamak\,\cite{Hollmann2025}. They have been successfully modeled by a novel multi-physics work-flow capable of describing the thermomechanical response of brittle sublimating PFCs, initially focusing on the description of the onset of brittle failure\,\cite{Ratynskaia_2025}, but soon expanded to the description of the fragmentation process and even to predictions of the solid debris speeds\,\cite{ITPA_Rizzi, EPS_Ratynskaia}. This work-flow has also provided a satisfactory description of accidental RE-induced damage on boron-nitride tiles in WEST\,\cite{Rizzi_2025a}. 

To correctly describe the thermomechanical response of ductile (at high temperature) materials which possess a stable liquid phase, such as tungsten (W), the material response code block of the existing work-flow must be extended to include a multiphase equation of state and viscoplastic constitutive models, which is the subject of ongoing work\,\cite{Ratynskaia_2025b}. The first controlled experiment on RE-induced damage on W tiles, which was performed in April 2025 in the WEST tokamak and documented a PFC explosion followed by the release of a debris cloud\,\cite{ITPA_Corre}, should provide adequate empirical constraints for future model validation. However, even within the simplification of pure thermal modeling, the vaporization induced, non-monotonic in-depth temperature profile, considered to be a trigger for the PFC explosion,\cite{De_Angeli_2023, Ratynskaia_2025, Ratynskaia_2025b}, has also been confirmed in Geant4-MEMENTO modeling of accidental W tile damage in WEST\,\cite{EPS_Ratynskaia}. 

Here this work-flow is systematically applied for predictive modeling of the thermal response of the ITER tungsten first wall (FW) armour to various RE incidence scenarios. The one-way coupled work-flow involves DINA\,\cite{Pitts_2025} and JOREK\,\cite{Bergström_2024} simulations of the RE wall loading, Geant4\,\cite{Allison_2016} Monte Carlo (MC) simulations of the RE volumetric energy deposition and MEMENTO\,\cite{Paschalidis2024, Ratynskaia2022_1} heat transfer simulations of the PFC thermal response. The focus lies on the short time response, which dictates the PFC damage through vaporization losses and melt production, and on the long time response, which determines the coolant interface temperature whose knowledge is critical for the assessment of the occurrence of tube rupture and water leaks.  

This assessment has been performed in support of the 2024 ITER re-baseline in which the original beryllium (Be) main chamber armour is entirely replaced by W\,\cite{Pitts_2025}. A modification of this magnitude requires a re-design of the FW and, in particular, detailed consideration of the consequences of RE impact given the very different material properties of W versus Be. Until the establishment of the code work-flow discussed here, such calculations have not been possible. The ITER re-baseline is thus an opportunity to address the important issue of FW armour resilience to RE events, both in terms of macroscopic damage to the armour material itself and to the bond interface at the cooling substrate for the final, actively cooled ITER FW to be installed prior to the start of DT operations. A key design question is the optimum thickness of the W armour in those FW locations where RE impact is most likely. More specifically, the work described here seeks to recommend a thickness which would at least permit a small number of worst case RE interaction events to be tolerated before replacement of damaged components is required.

The paper is organized in the following manner: Section \ref{sec:loading} reports the RE loading conditions provided by DINA \& JOREK simulations. Section \ref{sec:work-flow} discusses numerical aspects arising from the incorporation of the loading scenarios in the existing work-flow. Section \ref{sec:grazing} discusses general physical aspects of the RE volumetric heating and the PFC thermal response in ITER-relevant scenarios. Sections \ref{sec:results_DINA}-\ref{sec:results_JOREK} analyze the damage and thermal response results for the DINA and JOREK scenarios, respectively. Section \ref{sec:secondary_products} analyzes the characteristics of the secondary particles produced in the interaction cascade induced by RE incidence. A critical assessment of the results is provided in Section \ref{sec:assessment} and the conclusions can be found in Section \ref{sec:conclusions}.

\section{Loading specification}\label{sec:loading}

In the 2024 ITER re-baseline, two variants of the FW will be installed [1]: an inertially cooled, temporary FW for the Start of Research Operations (SRO) Phase, followed by a much more complex, actively cooled system for the DT Operation Phase. The rationale for a first, inertial wall, is precisely to allow margin for uncontrolled transients, particularly REs, when pushing the machine to its highest plasma current and field (in non-active phase). By the time of installation of the final, actively cooled variant, the expectation is that plasma operational and disruption mitigation experience will have reduced severe transient impacts to the very minimum and this final wall need only thus be designed to allow for a very small number of events. 

The new actively cooled W FW is currently still in the design phase, but it will probably be largely based on the technology already prototyped for the 2016 Baseline with Be armour\,\cite{Raffray2014}. This consists of two types of FW panel, Enhanced Heat Flux (EHF) and Normal Heat Flux (NHF), with the former, rated to $4.7\,$MW/m$^2$ stationary power handling capacity, incorporating hypervapotron cooling and the latter, rated to $2\,$MW/m$^2$ power handling capacity based on standard pipe cooling. Both variants comprise stacks of toroidally shaped fingers to form the full panel. In the EHF design, the toroidal fingers consist of a castellated armour layer, with the castellations such that the finger effectively comprises a chain of small "tiles" with dimensions (for the Be variant) of approximately $16\times16\,$mm$^2$. In the NHF case, the design uses much larger armour units. 

\begin{figure*}
    \centering
    \raisebox{-.5\height}{%
        \subfloat{%
            \begin{overpic}[width=0.28\linewidth]{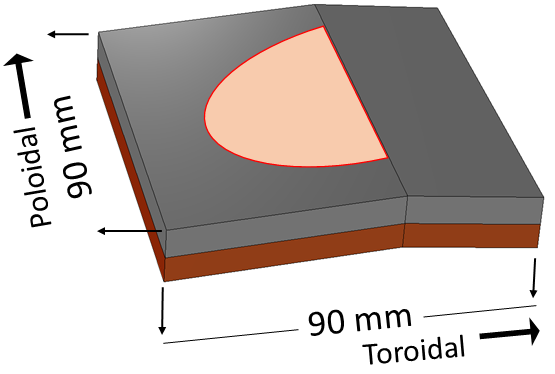}
                \put(-6,65){{\textbf{\big(a)}}}
            \end{overpic}
        }%
    }\hspace{0.5em}
    \raisebox{-.5\height}{%
        \subfloat{%
            \begin{overpic}[width=0.31\linewidth]{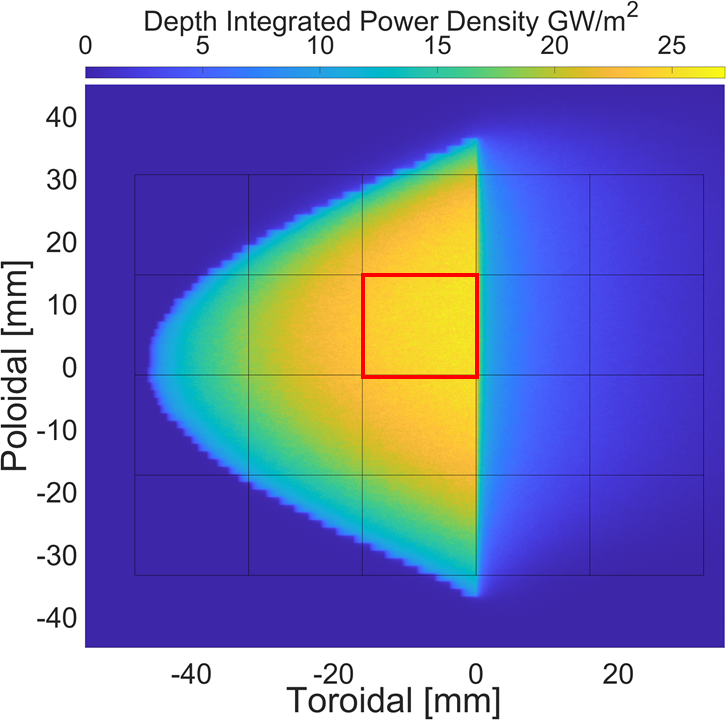}
                \put(-6,70){{\textbf{\big(b)}}}
            \end{overpic}
        }%
    }\hspace{0.5em}
    \raisebox{-.5\height}{%
        \subfloat{%
        \begin{overpic}[width=0.15\linewidth]{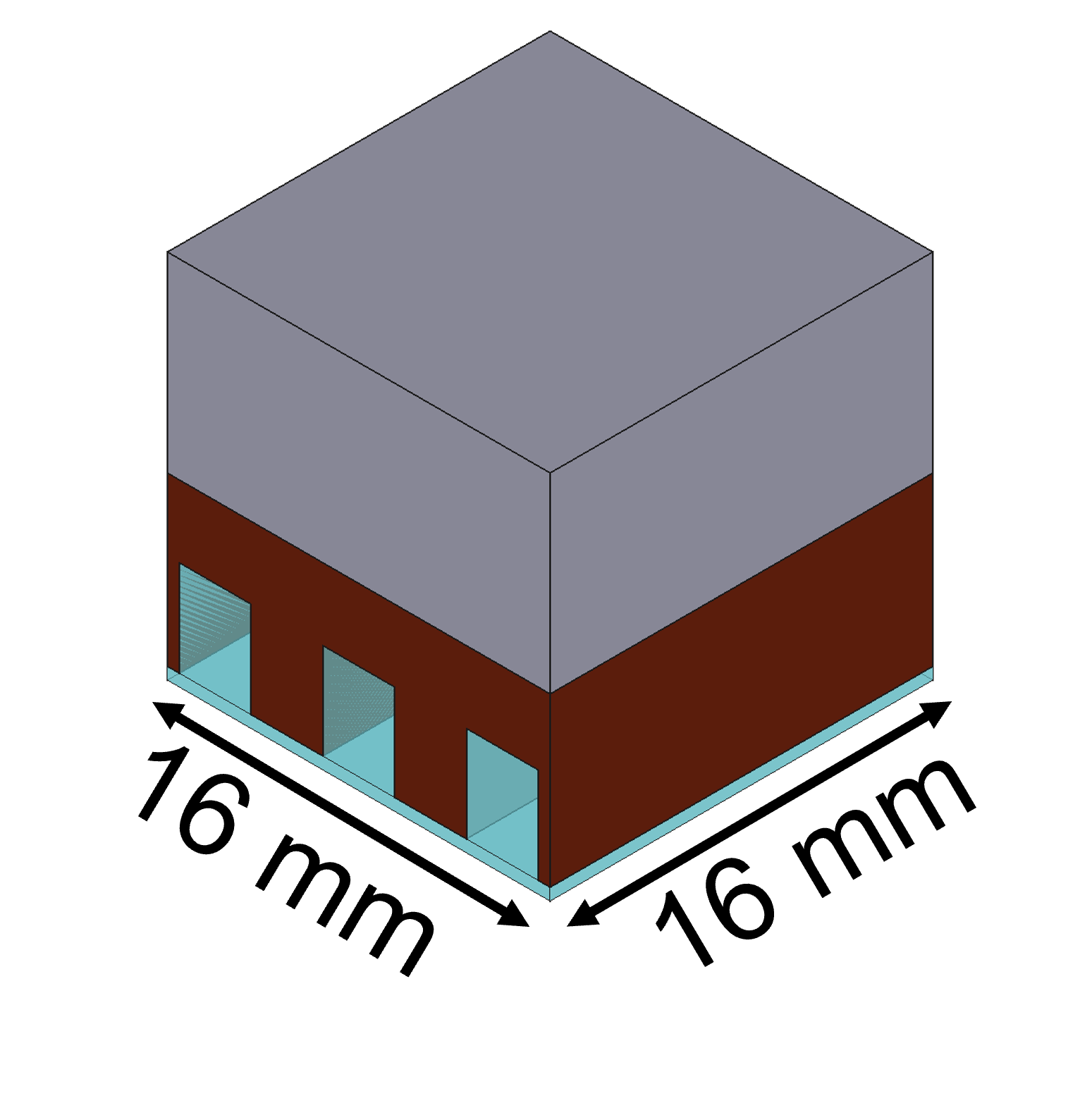}
                \put(-6,95){{\textbf{\big(c)}}}
            \end{overpic}
        }%
    }\hspace{0.5em}
    \raisebox{-.5\height}{%
        \subfloat{%
            \begin{overpic}[width=0.15\linewidth]{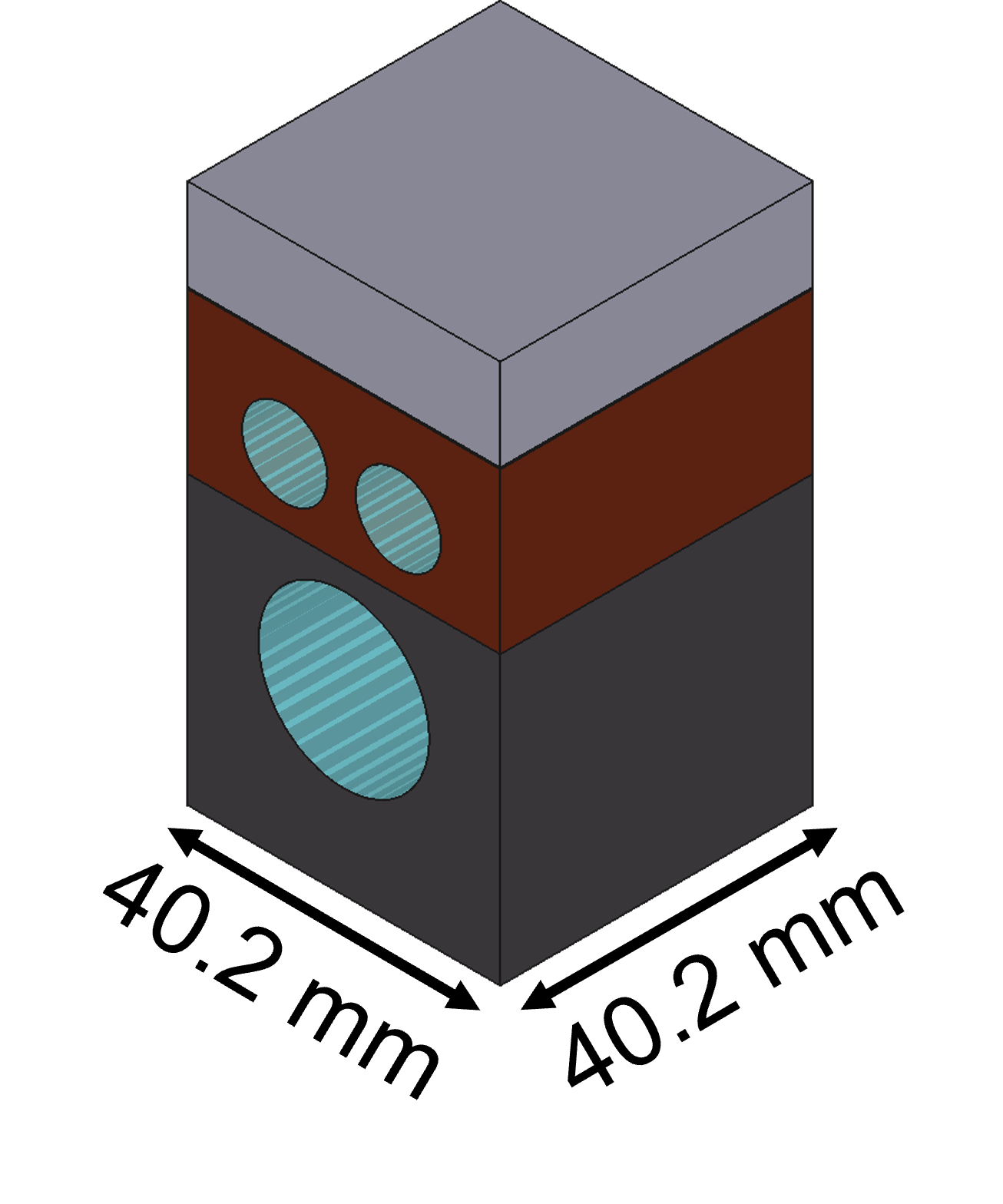}
                \put(-15,85){{\textbf{\big(d)}}}
            \end{overpic}
        }%
    }
    \caption{Numerical implementation of the problem. (a) 3D sketch of the apex simulated in Geant4 in all DINA scenarios, with the wetted area marked in pink. (b) Depth-integrated power density for the DINA scenario with 50\,kJ deposited over 1\,ms. The apex corresponds to the 0 mm position toroidally. The thin black lines indicate the castellations with the most loaded tile demarcated by a solid red line. (c) A single tile of an EHF panel, as simulated in MEMENTO in all DINA scenarios and the JOREK D1 scenario, made up of W and 7\,mm of CuCrZr. The different material layers and the hypervapotron position and shape can be clearly seen. (d) A single tile of an NHF panel, as simulated in MEMENTO for the JOREK D2 scenario, made up of W, 20.8\,mm of CuCrZr and 36.9\,mm of stainless steel. The different material layers and the cooling pipe position and shape can be discerned.}
    \label{fig:panel_implementations}
\end{figure*}

The majority of the RE impacts, are expected to occur in FW regions fitted with EHF panels, but some interactions on NHF units cannot be excluded. Since the new FW design is not yet available, here we assume representative "tile" sizes for the RE impact calculations of $16\times16$\,mm$^2$ and $40\times40$\,mm$^2$, as in the 2016 design (see also Fig.\ref{fig:panel_implementations}). The original Be armour had standard thickness of 8\,mm. In the present study, thickness is the key parameter we aim to study and so it is varied in the calculations between 8 and 15\,mm.  In an actively cooled system, and especially one made of W, there are of course restrictions on the thickness from the point of view of stationary power handling and the final choice will always be a compromise between the need for heat exhaust and transient resilience.

In the following, it is important to differentiate between the incident, deposited and absorbed energies. The \textit{incident} energy is the kinetic energy of the RE beams as they wet the surfaces of interest. This is generally larger than the \textit{deposited} energy (which is employed as the volumetric heat source in the temperature equation) due to the backscattering or transmission of primary electrons and secondary particles. Finally, the \textit{absorbed} energy is also a fraction of the deposited energy due to energy losses from the surface processes, which are dominated by vaporization losses in the regimes of interest.

\subsection{DINA scenario}

Considering a pre-disruption plasma current of 15 MA, we use a simulated DINA-DS scenario which predicts the formation of a 9\,MA RE beam during the current quench of an upward-going vertical displacement event (VDE)\,\cite{Pitts_2025}. For the RE termination and associated loading analysis, a magnetic equilibrium at the time when $q_{95} = 2$ is selected. At this safety factor, the destabilization of magnetohydrodynamic (MHD) instabilities (which is not, of course, simulated by DINA) is expected to trigger a sudden RE loss, with the electrons predominantly impacting the top of the main chamber, in particular at FW panel $\#$8. The RE footprint on the panel is then calculated by the SMITER 3D field line tracing code assuming a beam width of $\Delta_{RE}=4$ mm, based on the Larmor radius of the relativistic electrons. The scenario is depicted in Fig.16 of Ref.\cite{Pitts_2025}. 

Further assumptions comprise: \textbf{(i)} The RE energy distribution follows an exponential dependence \cite{Rosenbluth_1997}, $\propto\exp(-E/E_0)$, with $E_0=15$\,MeV that is truncated at 1\,MeV and 50\,MeV; \textbf{(ii)} all REs are incident at 5$^{\circ}$ with respect to the surface; \textbf{(iii)} the toroidal magnetic field of 7\,T at the panel location is also inclined at 5$^{\circ}$ with respect to the surface, with a direction away from the panel. This translates to a zero pitch angle; \textbf{(iv)} two energy deposition durations are probed, 1\,ms and 100\,ms. The former corresponds to a case where REs only deposit their kinetic energy right after formation, while the latter represents a case where the RE beam is preserved for a longer time thus allowing for conversion of magnetic energy in the RE beam to RE kinetic energy\,\cite{MartinSolis2025}; \textbf{(v)} the deposited energy is scanned up to the maximum energy possible in the simulations, which is dictated by the validity of the W thermo-physical properties (see Sections \ref{sec:grazing} $\&$ \ref{sec:assessment} for further details). We note that in the energy scans we are essentially just varying the number of incident REs. 

\subsection{JOREK scenarios}

The JOREK simulation scenarios are summarized in Table \ref{tab:JOREKtab}. They intend to explore various cases of RE beam termination on ITER, in particular investigating the possible benefits of a second deuterium injection into an established RE beam by varying plasma resistivity and ion density during the termination phase. The 4 scenarios (designated as D1-D4 in Table \ref{tab:JOREKtab}) consider different cases, with three "unmitigated" simulations (D1, D3, D4) and a fourth "mitigated" case (D2). The load estimates are derived through post-processing of preceding disruption simulations in JOREK using a fluid model for the REs which is self-consistently coupled to the MHD model. The simulations, described in Refs.\cite{Bandaru_2025,Bandaru_2024}, consider a high current $14.6\, \rm MA$ scenario. The equilibrium, pre-disruptive plasma is first subjected to an "artificial thermal quench" with current profile flattening. This is followed by a spatially uniform introduction of deuterium and neon and a small RE seed (0.1\,A) which is then subject to avalanching in the current quench plasma. At the end of the current quench phase, the RE current $I_{RE}$ is roughly $9\, \rm MA$, similar to the DINA scenario.

In common with the DINA simulation, the current begins to be scraped-off at the boundary and partly re-induced in the core, leading to a decrease in $q_{95}$. Up to this point, the simulation is performed in an axisymmetric (2D) configuration. Once the edge safety factor reaches $q_{95}\approx2.2$, the simulation is continued in 3D, and the prompt growth of a tearing mode instability is observed. The ``unmitigated'' cases used the temperature dependent Spitzer resistivity $\sim 10^{-4}\, \rm \Omega m$, whereas for the ``mitigated'' scenario D2, the resistivity was increased by a factor 10 leading to a more violent instability. In all cases, a near complete loss of the full RE current was observed.

To determine the RE deposition during the MHD event, full-orbit particle tracers were initialized at the onset of the termination up to collisions with the FW (which is modelled in 3D, including the baseline FW panel shaping design for the final, actively cooled wall). The analysis here concerns the most loaded panel identified in each scenario. Different energy distributions for the REs were used for the particle tracing. The simplest case assumes a mono-energetic distribution, where the total energy is estimated by integrating the acceleration from the electric field in time, accounting for energy losses caused by collisions and Bremsstrahlung, yielding a total kinetic energy of about $24\, \rm MJ$ (cases D1 and D2). A small pitch angle is assumed for the particles, such that $p_\parallel/p_{\rm total} = 0.99$, where $p_\parallel$ is the particle momentum along the magnetic field line and $p_{\rm total}$ the total particle momentum (cases D1 and D2). In addition, a more realistic avalanche distribution given analytically in Ref.\cite{Embreus_2018} is employed in cases D3-D4. Owing to the strong parallel electric field, this also yields very small pitch angles in the range of $0.99$-$1$. Finally, to evaluate the effect of higher pitch angles, a modified version of the avalanche distribution is used, where the exponent is altered to make the distribution much broader. The full details of the particle tracing can be found in Ref.\cite{Bergström_2024}.

\begin{table*}[!h]
\centering
\renewcommand{\arraystretch}{1.2}
\begin{tabular}{m{4.9cm} m{2.cm} m{2.3cm} m{3.0cm} m{2.8cm} }
\hline

\textbf{Scenario ID} & \textbf{D1} &\textbf{D2} & \textbf{D3} & \textbf{D4} \\
\hline
\textbf{RE pitch angle [rad]} & 0.99 & 0.99 & 0.99--1 & $\exp(-10 \times \text{pitch})$ \\
 
\textbf{RE energy distribution} & \shortstack[l]{Monoenerg.\\26 MeV} & \shortstack[l]{Monoenerg.\\26 MeV} & \shortstack[l]{Avalanching distr.\\(mean 2 MeV)} & \shortstack[l]{Avalanching distr.\\(mean 2 MeV)} \\

\textbf{RE MHD termination type} & Unmitigated & Mitigated & Unmitigated & Unmitigated \\

\textbf{Interaction time [ms]} &  1, in bursts & 0.25, in bursts &  1, in bursts &  1, in bursts \\

\textbf{Energy per panel [MJ]} & 1.2 & 2.19  & 1.43 & 1.12 \\

 \rowcolor{gray!20}
\textbf{Max energy per tile [kJ]} & 10.5 & \textbf{\textcolor{red}{40*}}  & 16.5 & 17 \\
  \rowcolor{gray!20}
\textbf{Max energy density [J/cm$^3$]} & 5.6$\times$10$^4$ &3.8 $\times$10$^4$   & 8.1 $\times$10$^5$ & 8$\times$10$^5$ \\
  \rowcolor{gray!20}
\textbf{$\boldsymbol{T^\mathrm{surf}_\mathrm{max}}$ [K]  /  $\boldsymbol{T_\mathrm{max}}$ [K]} & 6760/7970 &  6450/6500   & N/A & N/A \\
  \rowcolor{gray!20}
\textbf{$\boldsymbol{h_\mathrm{melt}}$ [$\boldsymbol{\mu}$m]} & 880 &  500   & N/A & N/A \\
  \rowcolor{gray!20}
\textbf{$\boldsymbol{h_\mathrm{vap}}$ [$\boldsymbol{\mu}$m]} & 25 & $\sim$0.1   & N/A & N/A \\
\hline
\end{tabular}
\caption{Summary of simulations with JOREK input. The first five rows describe the JOREK scenarios following the notations of Ref.\cite{Bergström_2024}: scenario ID, pitch angle and energy distribution as well as MHD scenario specification, RE-PFC interaction time and incident energy per most loaded panel. The following rows highlighted in gray summarize the Geant4 simulation results (energy deposited into the most loaded tile and maximum energy density), followed by the MEMENTO simulation results (maximum surface temperature and maximum bulk temperature values as well as melt and vaporization losses). Note that the maximum energy per tile for the D2 scenario (in red) stands out due to more than 6 times larger tile area in that case.}
\label{tab:JOREKtab}
\end{table*}

\begin{figure*}[!h]
    \centering
    \addtolength{\tabcolsep}{-9pt} 
    \begin{tabular}{cccc}
        \subfloat{%
          \begin{overpic}[width = 1.66in]{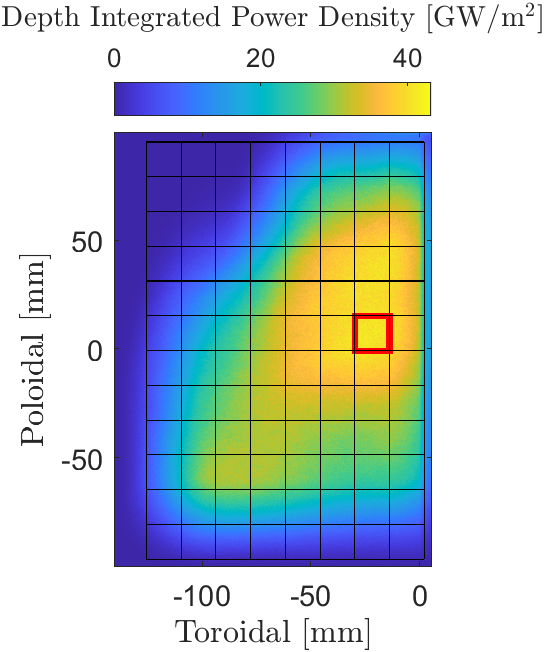}
          \put(2,76){\hbox{\kern3pt\textcolor{black}{\textbf{D1} }}}
          \end{overpic}
        }&
        \subfloat{%
          \begin{overpic}[width = 1.7in]{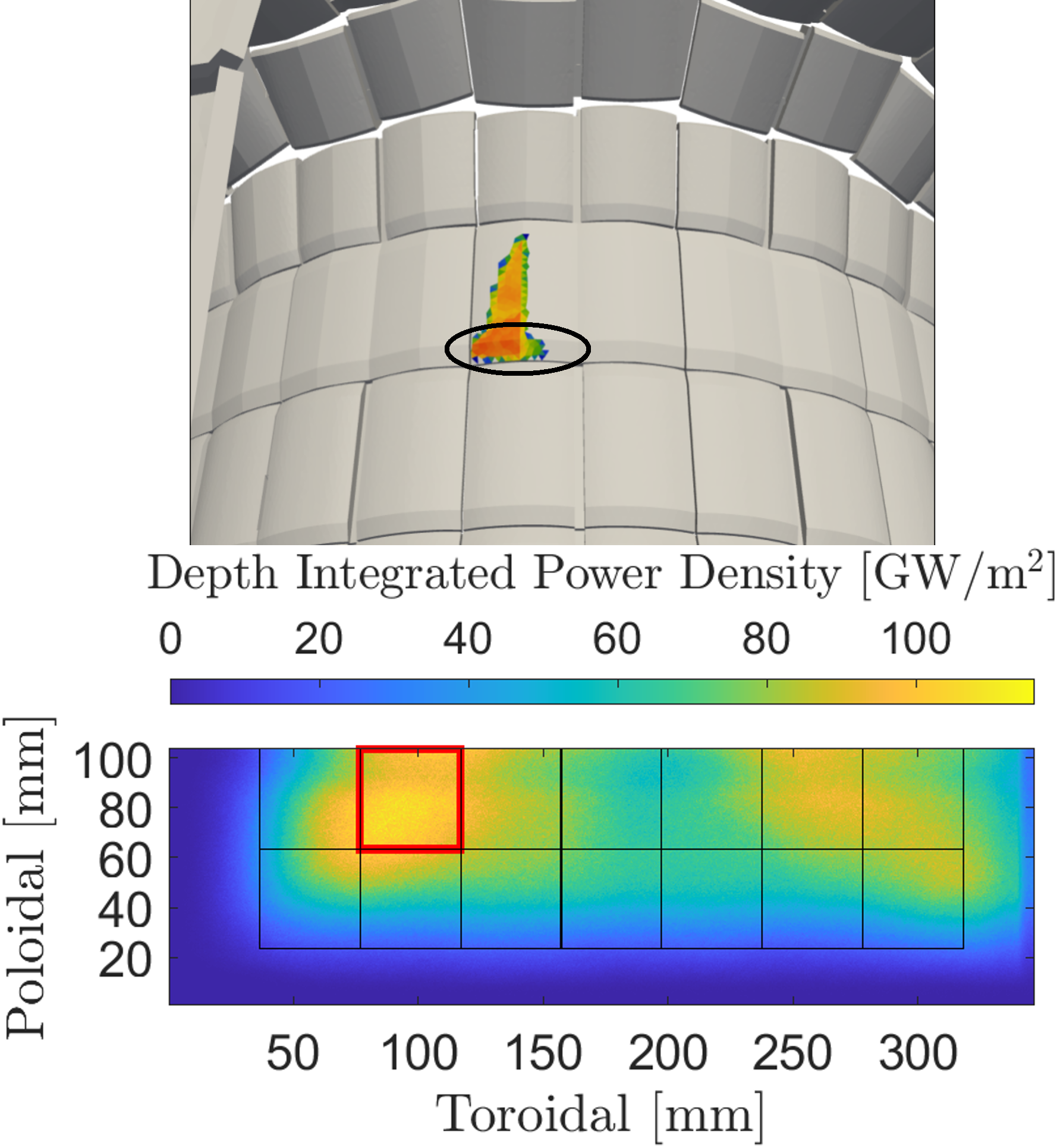}
          \put(1,81){\hbox{\kern3pt\textcolor{black}{\textbf{D2} }}}
          \end{overpic}
        }
        \subfloat{%
          \begin{overpic}[width = 1.66in]{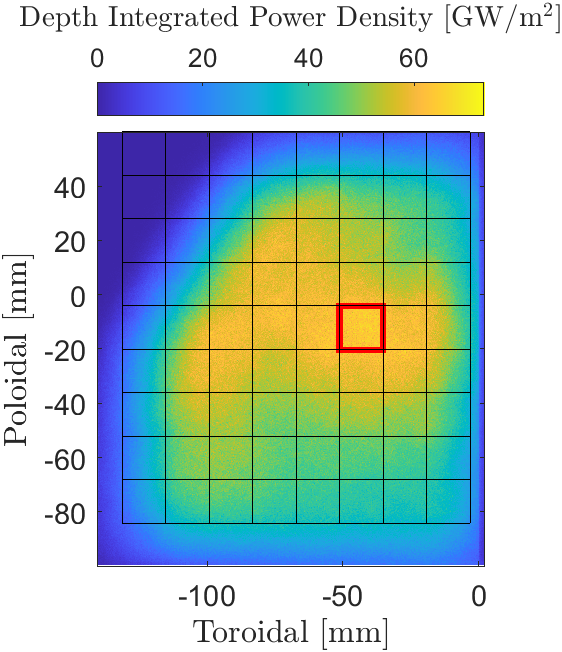}
          \put(0,79){\hbox{\kern3pt\textcolor{black}{\textbf{D3} }}}
          \end{overpic}
        }&
        \subfloat{%
          \begin{overpic}[width = 1.7in]{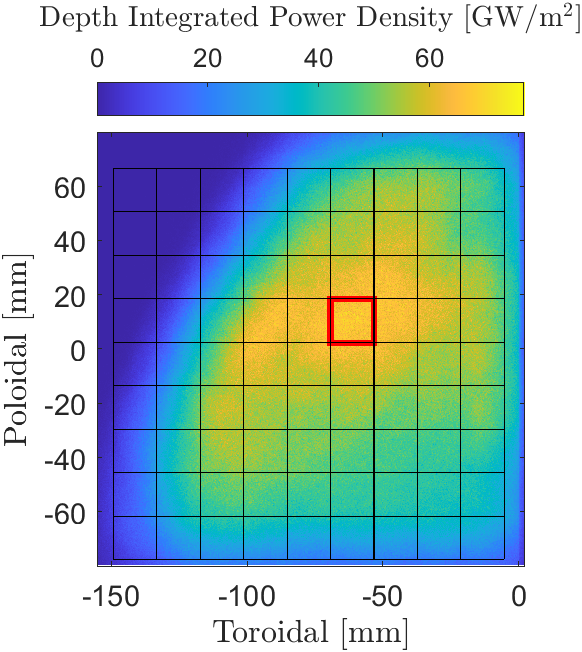}
          \put(0,78){\hbox{\kern3pt\textcolor{black}{\textbf{D4} }}}
          \end{overpic}
        }\\
        \subfloat{%
          \begin{overpic}[width = 1.7in]{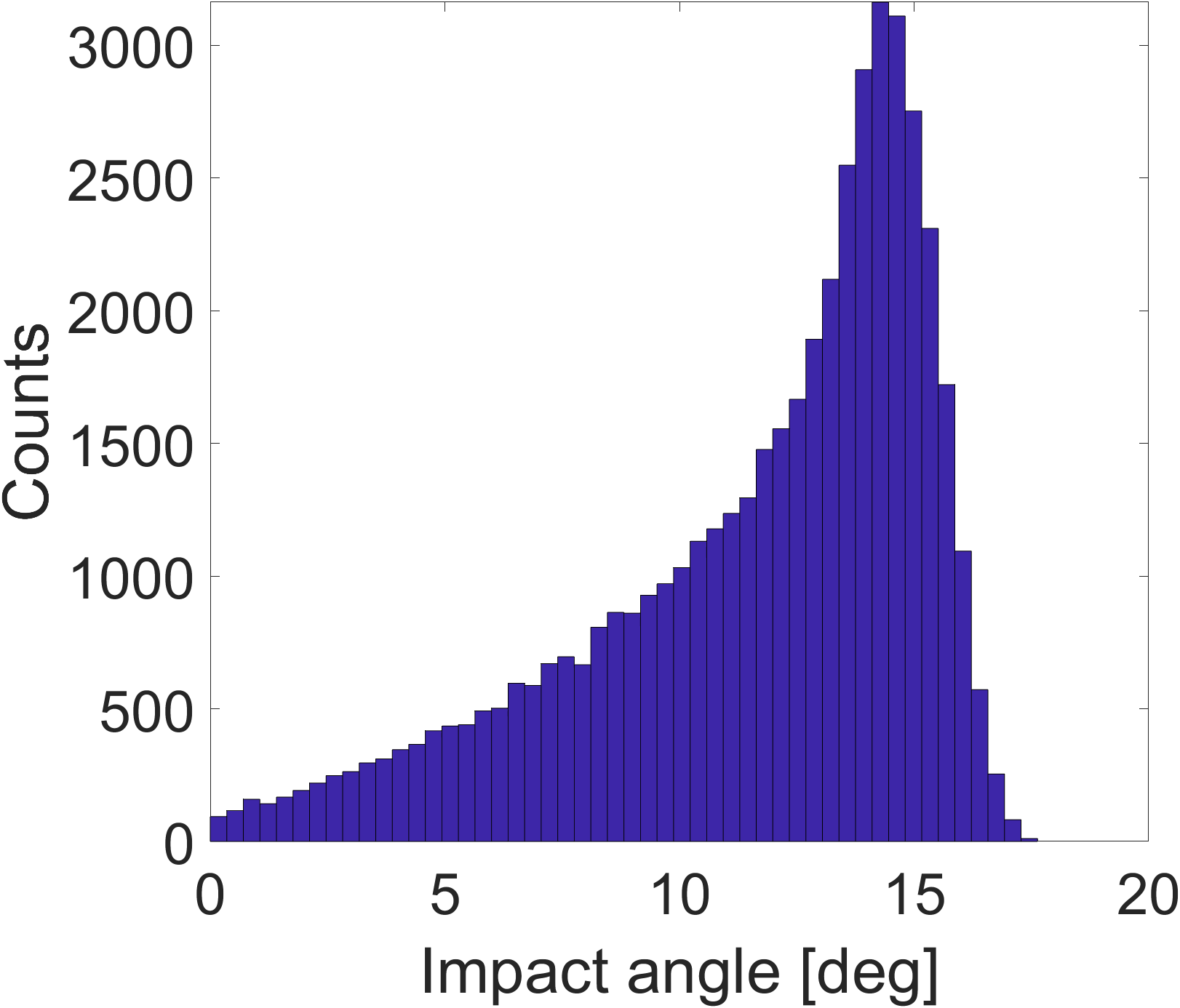}
          \put(20,75){\hbox{\kern3pt\textcolor{black}{\textbf{D1} }}}
          \end{overpic}
        }&
        \subfloat{%
          \begin{overpic}[width = 1.7in]{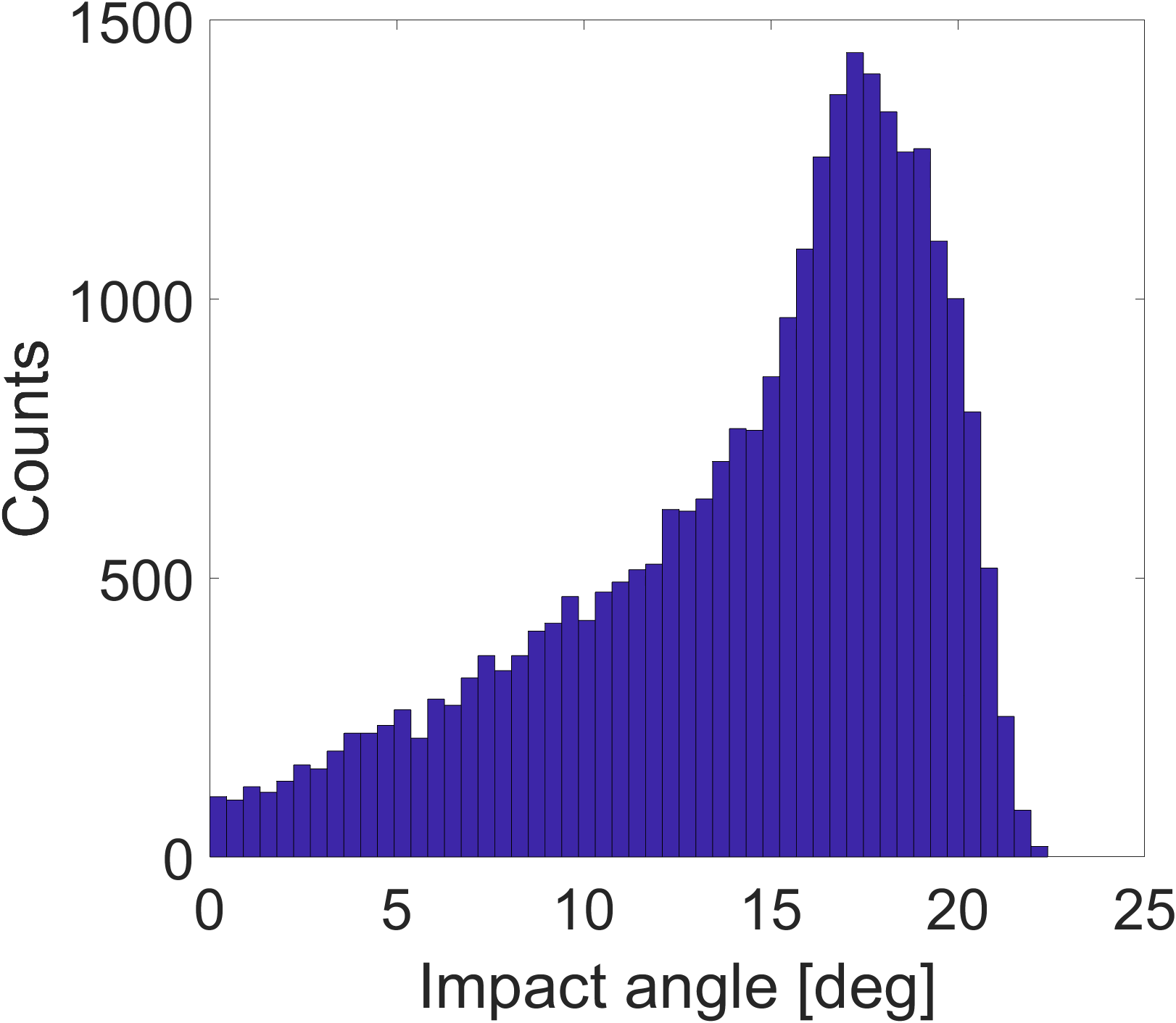}
          \put(20,75){\hbox{\kern3pt\textcolor{black}{\textbf{D2} }}}
          \end{overpic}
        }
        \subfloat{%
          \begin{overpic}[width = 1.7in]{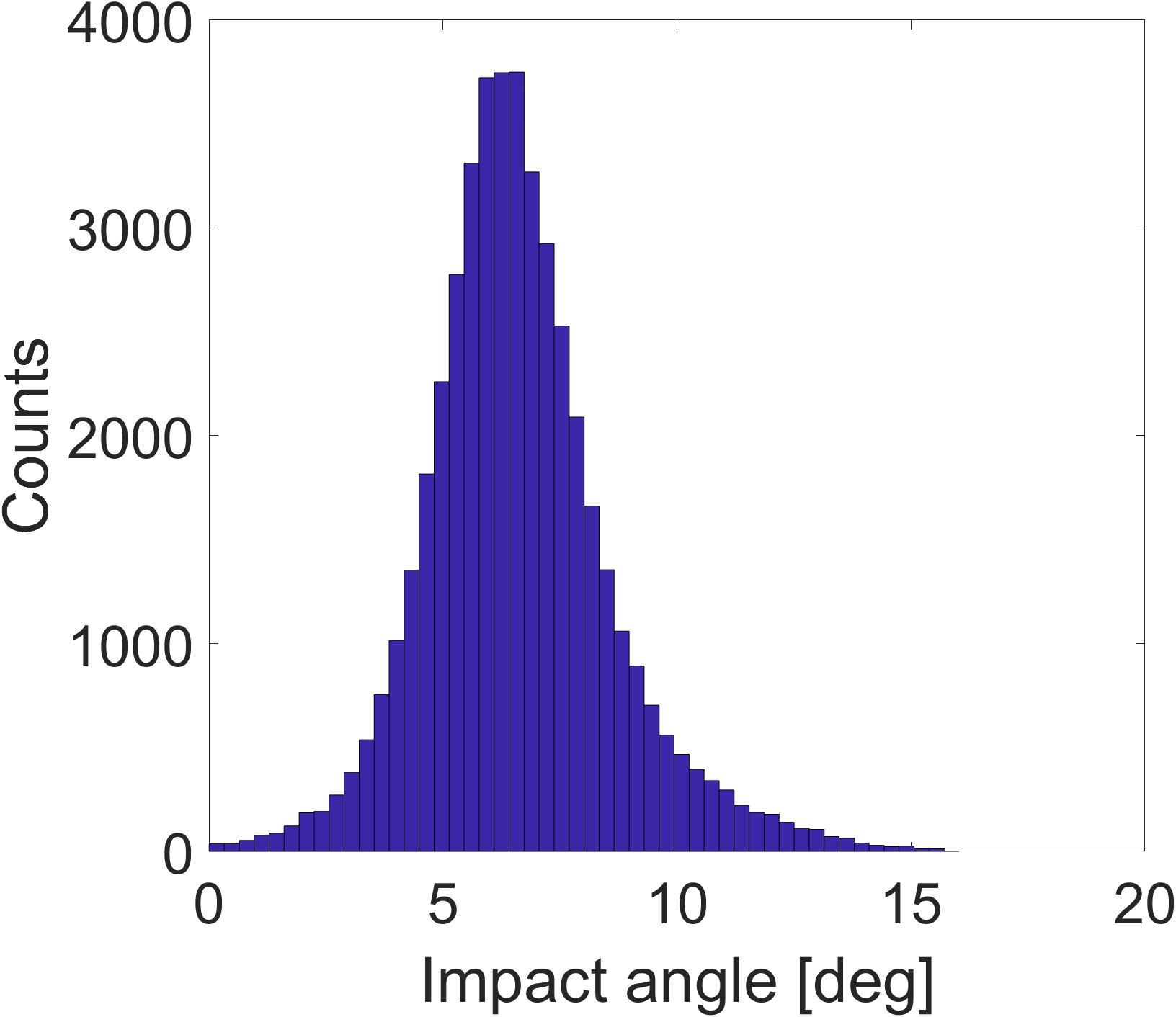}
          \put(20,75){\hbox{\kern3pt\textcolor{black}{\textbf{D3} }}}
          \end{overpic}
        }&
        \subfloat{%
          \begin{overpic}[width = 1.7in]{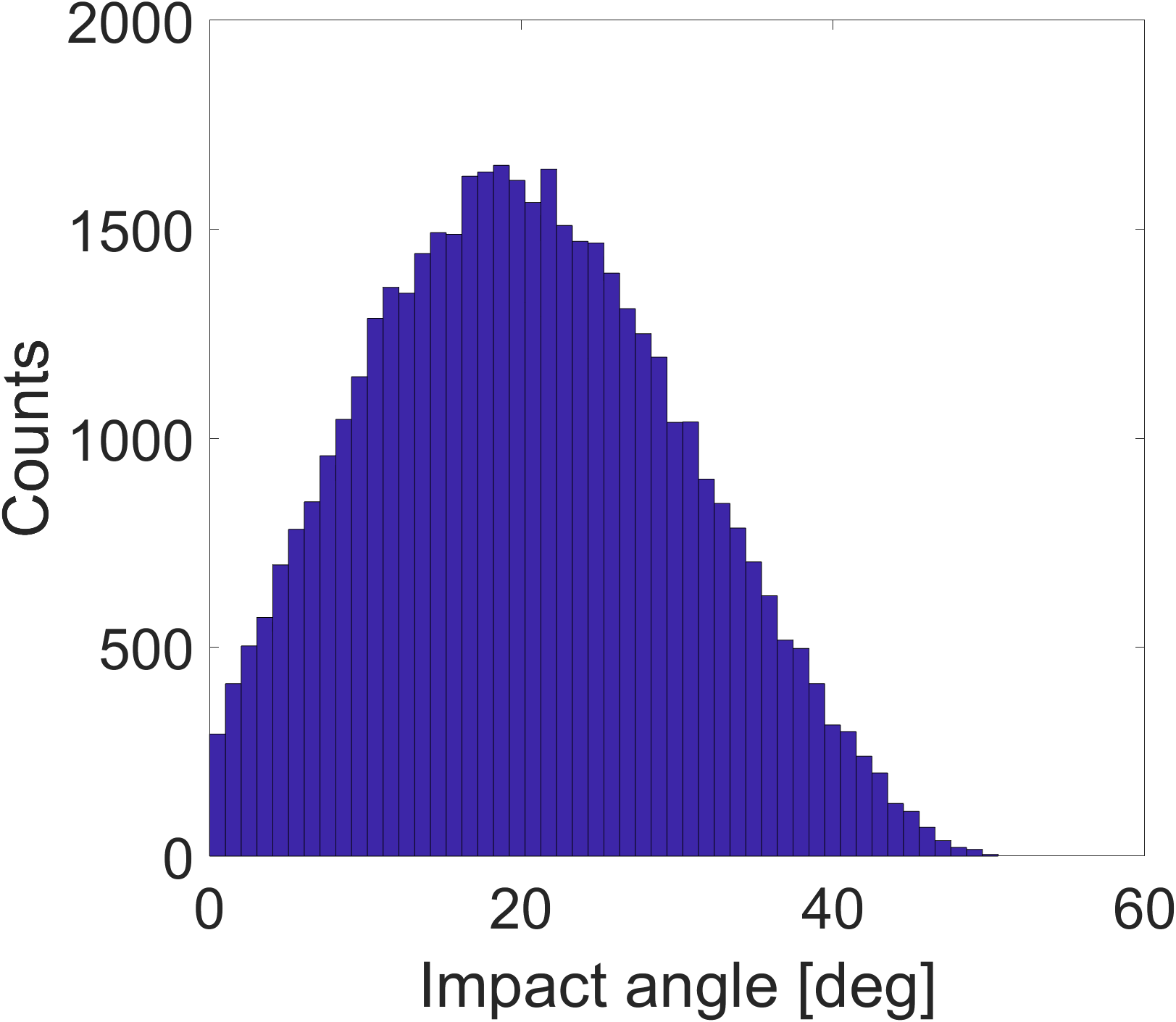}
          \put(20,75){\hbox{\kern3pt\textcolor{black}{\textbf{D4} }}}
          \end{overpic}
        }\\
    \end{tabular}
    \addtolength{\tabcolsep}{9pt} 
    \caption{Results for the JOREK scenarios D1-D4 summarized in Table \ref{tab:JOREKtab}. Geant4 calculations of the depth-integrated power density profiles (upper row). Histograms of the impact angles in each scenario (lower row). Note that the heaviest loaded region in the D2 scenario falls on the sloped part of the panel, see the circled area in the gray picture.}
    \label{fig:JOREK_Geant4_results}
\end{figure*}

\section{The Geant4 - MEMENTO work-flow}\label{sec:work-flow}

\subsection{Modeling of the energy deposition}

\subsubsection{Simulation of RE transport and of the accompanying particle shower.}

Volumetric energy deposition is evaluated by MC simulations of RE transport inside a W armoured panel, here performed with the Geant4 open source software\,\cite{Geant4_2003,Allison_2006,Allison_2016}. Our Geant4 simulations of relativistic electron transport inside high-Z materials have been benchmarked against a wide dataset of calorimetry measurements for primary electron energies ranging from 0.1 to 1\,MeV\,\cite{Lockwood_1980}. Furthermore, the Geant4 simulations have also been benchmarked against measurements of normal and oblique keV electron backscattering from pure W\,\cite{Bronshtein_1969, Gomati_2008,Hunger_1979,Reimer_2000} as well as against measurements of relativistic MeV-range electron backscattering from other high-Z metals\,\cite{Tabata_1967}. These studies enabled the selection of a computationally efficient and accurate physics list that covers all electromagnetic processes relevant to primary RE electron transport and to secondary particle generation \& transport. The reported Geant4 simulations were carried out with the single scattering implementation available in the G4EmStandardPhysicsSS library. This library provides a highly accurate description of nuclear scattering through the G4eDPWACoulombScatteringModel that is based on Dirac partial wave analysis with the ELSEPA code\,\cite{Salvat_2021} and of all other elastic \&  inelastic electromagnetic processes through models from the PENELOPE\,\cite{Penelope} and Livermore\,\cite{Depaola_2006,Depaola_2003,Depaola_2000} libraries. Finally, neutron generation (primarily through photo-nuclear reactions) and subsequent transport are described with the aid of the G4NeutronHP model, using the ENDF/B-VII.0 cross-section library\,\cite{Mancusi_2014}.

\subsubsection{Set-up of the problem for the DINA scenario.}

The 3D panel geometry is faithfully reconstructed to ensure that generation of cascade products and transport of primary/secondary particles are accurately described. A $90\times90$\,mm$^2$ domain is simulated, encompassing about 20 single tiles, as depicted in Fig.\ref{fig:panel_implementations}(a). Panel loading with sufficient MC statistics and spatial resolution is not computationally feasible for the entire 3D domain. Hence, we have devised the following two-step approach. 

First, aiming to obtain a global picture that accounts for geometrical effects as well as for particle backscattering and transmission, the full 3D apex is loaded by a uniformly distributed particle source of $10^7$ electrons, allowing wetting of the specified area. The results are presented in 2D after depth-integration of the energy density map in Fig.\ref{fig:panel_implementations}(b). The loading is clearly nonuniform in spite of the uniform RE impact distribution due to the effect of the wetted area geometry (the extension decays along the toroidal direction), the geometry of the apex and the presence of the ambient magnetic field. The most loaded area, marked in the figure with a red line, is located just below the apex and is shifted with respect to the poloidal center of the wetted area in the direction of the backscattered electron gyration (see Sec.\ref{sec:grazing}).

The weak gradients over a single tile area, enable the second step where the energy deposition map of a single-point particle source with $10^6$ electrons is utilized to uniformly load a single $16\times16$\,mm$^2$ tile. This drastically enhances the particle statistics (reducing the MC uncertainty) and allows the exploration of a higher spatial resolution. Non-uniform meshing is employed in all Geant4 simulations, which allows the steep gradients in the near surface layers to be resolved. In the DINA scenarios, the upper layer was meshed with 20$\,\mu$m cells, relaxing gradually to 100\,$\mu$m cells in the bottom layer. Finer mesh sizes have also been probed, see Sec.\ref{sec:grazing} for details.  


Finally, the normalization for the single "worst case" tile, which is employed in further calculations, is deduced from the ratio of the energy deposited into the apex to the energy deposited over the most loaded region. For example, the case of "50\,kJ" corresponds to 6.7 kJ on the most loaded tile. This can be obviously rescaled for any other deposited energy values.

\subsubsection{Set-up of the problem for the JOREK scenarios.}

For each macroscopic marker, the JOREK simulations directly provide detailed impact characteristics on the FW surface, namely; the impact position and timing along with the incident momentum and local magnetic field. Each marker represents a fraction of the RE population, corresponding to a larger number of actual physical particles, quantified by a weight \(w_i\)\,\cite{Bergström_2024} which is used to reconstruct the total incident energy loaded.

The incorporation of the JOREK output in the Geant4 simulations is done in the following manner. Each marker is modeled  by a circular electron beam with a $20\,$mm radius, centered at the marker's impact position and retaining the marker momentum vector. The $20\,$mm radius matches the average size of the triangular mesh elements used in JOREK to describe the FW. Every source (every marker) is represented by \(10^5\) particles. Given that the number of markers per apex lies in the range $10^3-10^4$, this implies overall statistics of $10^8- 10^9$ particles. 

Within the wetted panel area, the magnetic field variations are between 4\% and 8\%.  Hence, a uniform magnetic field has been assumed in Geant4, with the average magnitude and inclination angle with respect to the FW panel, which correspond to 7.2\,T, $\sim5^{\circ}$ for scenarios D1, D3, D4, and to $8.2\,$T, $\sim11^{\circ}$ for D2.

In each scenario, several FW panels are impacted, but the focus lies on the most heavily loaded panel. Aiming to reduce the computational effort, only the most loaded region of the panel wetted area is included in the Geant4 simulations. The JOREK 3D panel geometry is reconstructed with the following simplification: the slightly curved surface is approximated as flat. Due to the large radius of curvature, this introduces $<3^\circ$ error in the marker impact angle, which is small given the impact angle distributions, see Fig.\ref{fig:JOREK_Geant4_results}.

For the high RE energy scenarios D1-D2, the first layer of 1.2\,mm is meshed with 50\,$\mu$m cells,  
gradually relaxing to 250\,$\mu$m cells in the bottom layer. For the D3-D4 scenarios with distributions dominated by low energy electrons, a finer resolution was used with a mesh of 20\,$\mu$m in the upper layers, relaxing to a 120\,$\mu$m mesh in the bottom layer.


Finally, the total energy incident on each panel, as specified from the JOREK results, is employed to re-normalize the energy maps produced by Geant4 to the deposited energy, after accounting for the losses due to the backscattering and transmission of particles, see Sec.\ref{sec:secondary_products} for more details.

\subsection{Modeling of the thermal response}

The Geant4 deposited energy density maps are converted to power densities, under the assumption of a uniform loading during the interaction time. These power densities constitute the volumetric heat source in the MEMENTO simulations. 

A worst case thermal response is considered on the single tile with the highest load, given that no heat diffusion can take place across the castellation gaps and that the contact with the stainless steel substrate through the CuCrZr layer is negligible on the time-scale of interest (the maximum temperature at the interface is reached after $\sim1$\,s).

The single tile of the EHF panel is modeled with different W layer thicknesses (8,10,12 and 15\,mm), a 7\,mm Cu-Cr-Zr layer and a simplified hypervapotron shape, where the teeth profile is aligned to the tile’s toroidal center, as shown in Fig.\ref{fig:panel_implementations}(c). The boundary conditions comprise; thermally insulated sides facing the gaps, vaporization cooling and thermal radiation cooling from the free surface and convection cooling flux at the CuCrZr - coolant interface. The employed heat transfer coefficient is a function of the wall temperature and corresponds to the case of an average water temperature of 343\,K, a 4\,MPa pressure and a 5\,m/s flow for the hypervapotron cooling system\,\cite{IOdoc_Raf}. Due to the uncertainties in the critical heat flux (CHF) under transient loading, values of 10\,MW/m$^2$\,\cite{IOdoc_Hirai} and 30\,MW/m$^2$\,\cite{Kim2022} are probed as the lower and upper thresholds of the expected CHF range. For the NHF panel tile, see Fig.\ref{fig:panel_implementations}(d), coolant velocities of 1.4\,m/s and 0.7\,m/s are assumed for the CuCrZr and stainless steel pipes, respectively,  while the water bulk temperature and pressure are set to 368\,K and 4\,MPa. The CHF is 10.5\,MW/m$^2$ for the CuCrZr pipe and 4.89\,MW/m$^2$ for the stainless steel pipe\,\cite{Tong1975}. 

The free surface is moving due to vaporization and the eroded cells are removed in the course of the MEMENTO simulation. This directly leads to a mismatch between the Geant4 and MEMENTO simulation domains. Since the GEANT4 simulations are based on the pristine (unaltered) geometry, Geant4 map loading without any translation to compensate for the evaporated material would lead to a loss of energy deposited at the upper cells.  The results reported here are calculated  with a shifted power density map along the radial direction, courtesy of the uniformity along the other two directions (valid for a single tile), to account for the fact that the free surface position is moving in time. The complex interplay between the evolving surface, the particles released and the incoming REs is discussed in Sec.\ref{sec:assessment}.

\section{Consequences of the grazing RE-incidence in the presence of a magnetic field} \label{sec:grazing}

\subsection{Energy deposition in the near-surface region}

\begin{figure}[!t]
\centering
\begin{overpic}[width=3.5in]{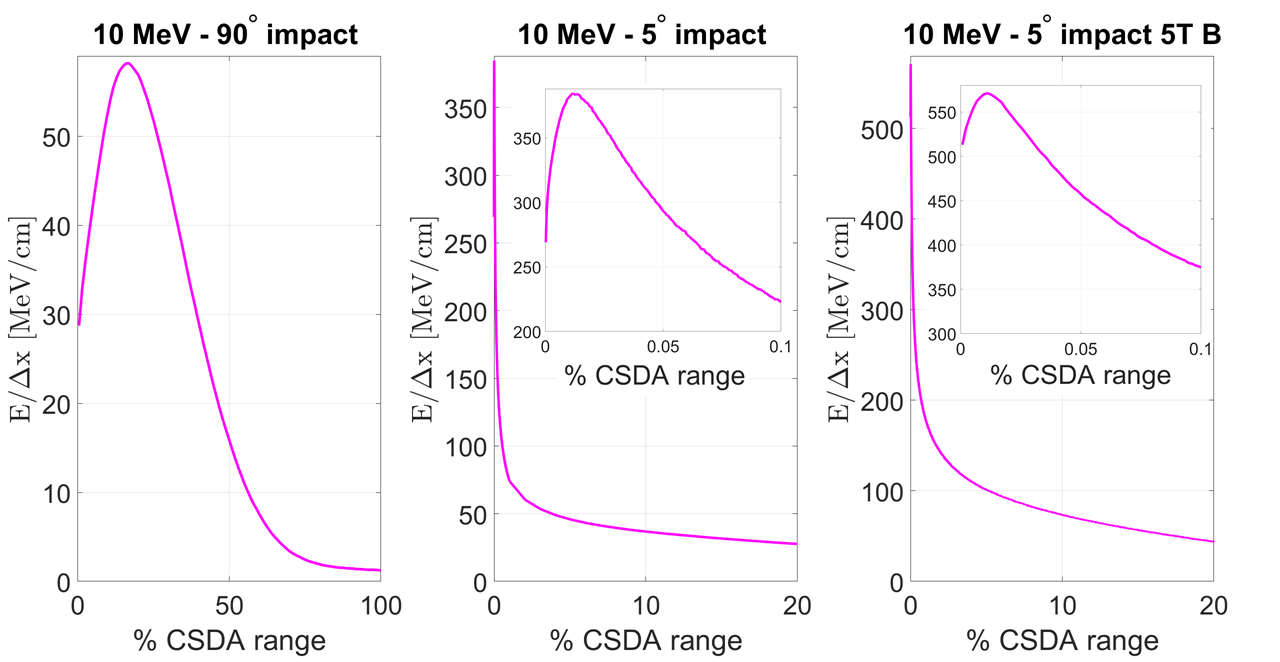}
    \put(6,8){\textbf{(a)} }
    \put(39,8){\textbf{(b)} }
    \put(72,8){\textbf{(c)} }
\end{overpic}
\caption{In-depth energy deposition profiles for impacts of 10 MeV REs on W at 90° (a), 5° (b), and 5° with B-field (c). Note that for 10 MeV and W, CSDA = 3.21 mm. }
\label{fig:impact_comparison}
\end{figure}


The continuous-slowing down approximation (CSDA) range is defined as the average path length traveled by incident energetic electrons in a given material until their complete thermalization\,\cite{Fano_1963}. The CSDA range values, which vary from $0.4\,$mm to $\sim8$\,mm in W for electron energies from 1\,MeV to 50\,MeV\,\cite{Berger_1992}, signify the volumetric nature of the energy deposition. However, these should not be viewed as representative energy deposition depths, since the energy deposition is non-local owing to the transport of energetic secondary products (mainly delta electrons and bremsstrahlung photons) and since the path length and the path depth always differ (especially in high-Z materials) due to the nuclear scattering. Moreover, in fusion relevant scenarios, the magnetic field can affect the charged particle trajectories, while the typically shallow impact angles magnify the difference between path length and path depth. 

In fact, at grazing angles, the electron path lies closer to the surface and the characteristic maximum of the energy deposition profile shifts towards the surface. Fig.\ref{fig:impact_comparison} shows an example for 10\,MeV RE energy, where the maximum moves from $\sim20\%$ of the CSDA range at normal incidence to $\sim0.01\%$ of the CSDA range at 5$^{\circ}$ grazing incidence or, correspondingly, from $0.65\,$mm to $300$\,nm.  Furthermore, while normal incidence electron backscattering is very limited in the MeV range, shallow impacts significantly enhance the backscattering yields. The average energy deposited in (a) is 8.81\,MeV per electron while in (b) it is $3.62\,$MeV. The energy losses are equally split between escaping photons and electrons in (a), while all extra losses in (b) are due to electron backscattering. 

The presence of B field outside the PFC leads to the re-deposition of backscattered electrons which has two consequences. First, as already known from the 1990s\,\cite{Bartels1994,Kunugi_1993}, it implies that more incident energy is absorbed. Indeed, the energy deposited in (c) increases to 9.04\,MeV. Note that, in order to accommodate the return of the majority of backscattered electrons, the domain size has to span a few Larmor radii in the perpendicular and a few helical steps along the B-field direction. Second, it can lead to non-uniform loading even with a uniform RE source, see Fig.\ref{fig:panel_implementations}(b). Moreover, despite the collision-dominated transport in high Z metals, the gyration of all charged particles due to B-field penetration inside the PFC has a finite effect on the energy deposition profile, as discerned by comparing the curves in Fig.\ref{fig:impact_comparison} (b), (c)  and the corresponding  profile maxima shown in the inserts. 
 

\begin{figure}[b]
    \centering
    \subfloat{%
        \includegraphics[width=5.1cm]{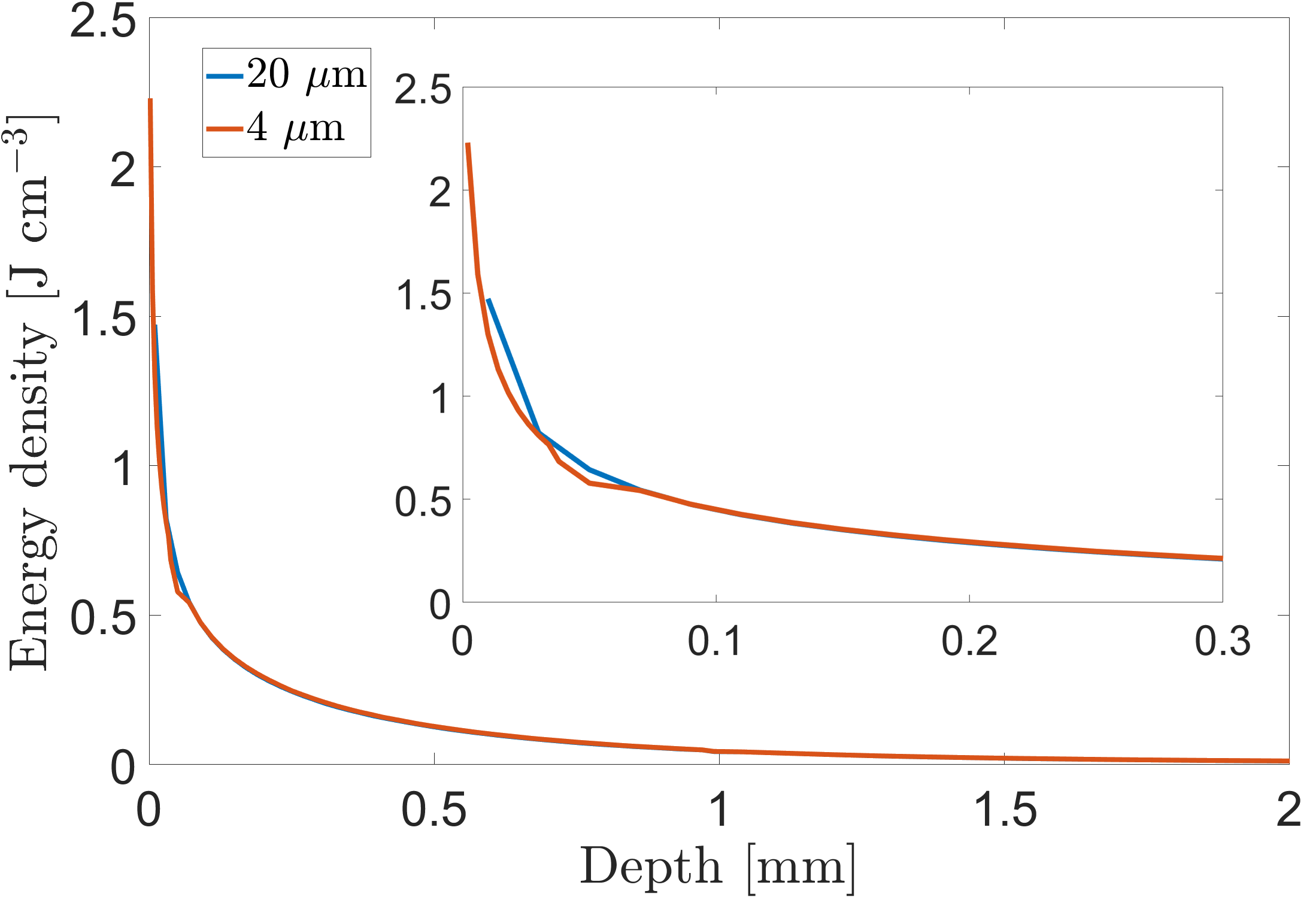}%
    }
    \hspace{0.5cm}
    \subfloat{%
        \includegraphics[width=5.1cm]{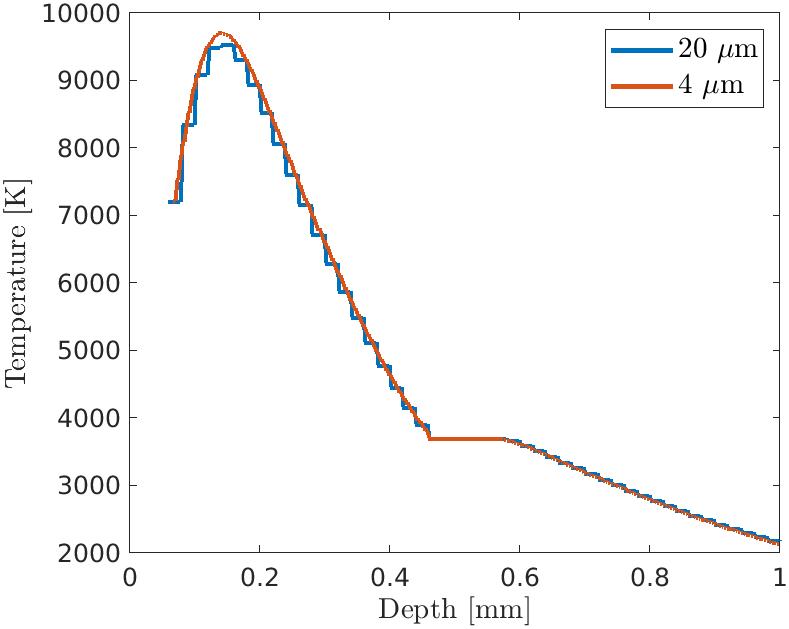}%
    }
    \caption{In-depth energy density profiles for the most loaded tile for the ‘50 kJ over 1 ms’ DINA scenario with an insert showing a zoom-in of the first 300 $\mu$m (a). The corresponding in-depth temperature profiles (b). The temperature profiles are shifted from the free surface (at 0 mm) due to erosion. Results are for resolutions of 4\,$\mu$m and 20\,$\mu$m.}
    \label{Resolution}
\end{figure}

Overall, these effects result in very shallow energy deposition with steep gradients that is confined to the upper few tens of micrometers. This necessitates high spatial resolution due to the exponential sensitivity of the vaporization to the surface temperature, which is dependent on the loading of the near-surface region. However, beyond a certain limit, the heat diffusion towards the top through the ultra thin layer should be sufficient to smooth these gradients. In this light, the sufficient discretization should be a fraction of the characteristic heat diffusion length during the RE loading time. Assuming the W diffusivity at the melting point\,\cite{Tolias2017_1}, the characteristic heat diffusion length is 120\,$\mu$m over 1\,ms implying a sufficient discretization of $\sim10\,\mu$m which will be by default adequate for longer loading times. This has been also verified in numerical tests with a spatial resolution down to 4\,$\mu$m. Results for the energy density simulated with $20\,\mu$m and $4\,\mu$m discretization are shown in Fig.\ref{Resolution}. They reveal appreciable, $\sim 30 \%$ differences in the upper cell value (see the insert). In fact, with better resolution even finer features can be found, see Fig.\ref{fig:impact_comparison} and the discussion above. Yet, as anticipated, the thermal response is identical when loading with the two volumetric heat maps of 20\,$\mu$m and 4\,$\mu$m resolution.

\subsection{Extreme temperatures and intense vaporization}

\begin{figure*}
    \addtolength{\tabcolsep}{-10pt} 
    \begin{tabular}{cc}
        \subfloat{%
          \begin{overpic}[width=0.25\linewidth]{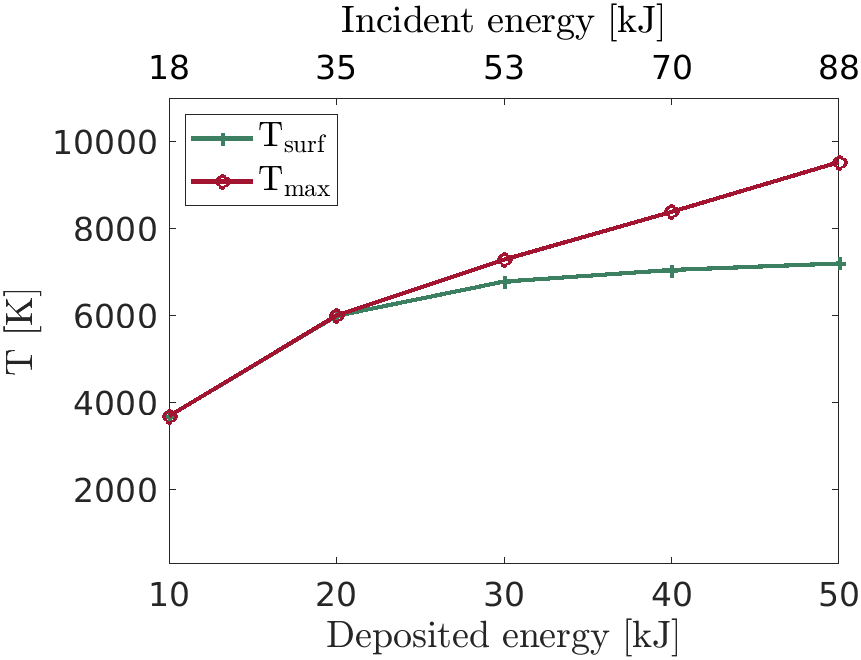}
            \put(25,15){{\textbf{\big(a)}}}
          \end{overpic}
        }&
        \subfloat{%
          \begin{overpic}[width=0.25\linewidth]{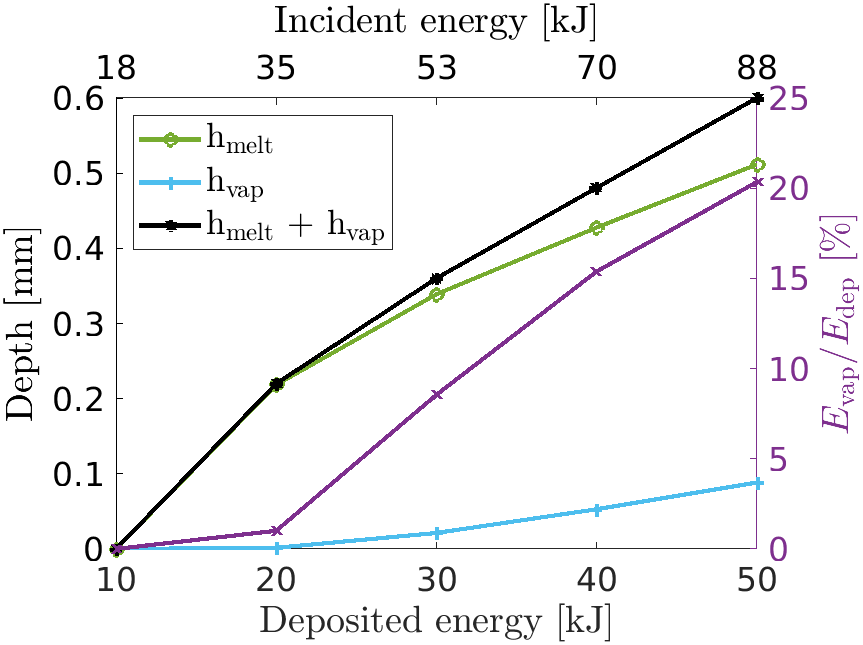}
           \put(70,20){{\textbf{\big(b)}}}
          \end{overpic}
        }
        \subfloat{%
          \begin{overpic}[width=0.25\linewidth]{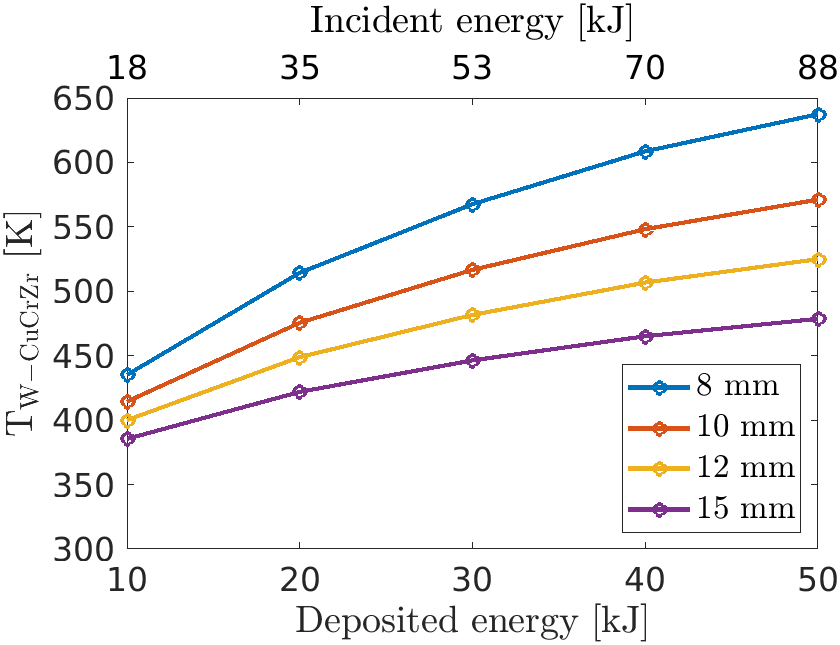}
          \put(25,15){{\textbf{\big(c)}}}
          \end{overpic}
        }
        \subfloat{%
          \begin{overpic}[width=0.22\linewidth]{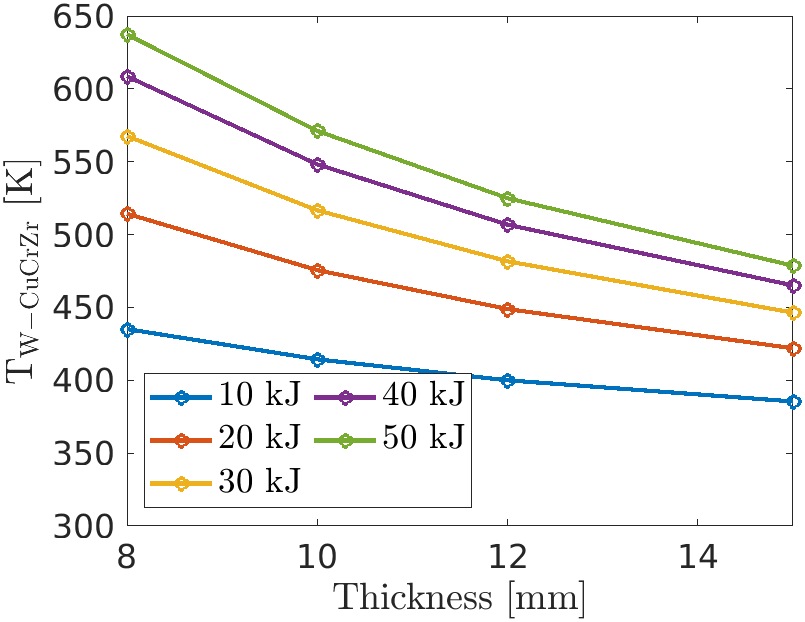}
            \put(85,14){{\textbf{\big(d)}}}
          \end{overpic}
        }
    \end{tabular}
    \addtolength{\tabcolsep}{6pt} 
    \caption{Surface damage (a-b) and cooling system response (c-d) in the DINA scenario for 1\,ms loading. The temperature at the W-CuCrZr interface is plotted versus the energy in (c) and versus the thickness in (d).}
    \label{fig:damage_1ms}
\end{figure*}

As evident from Fig.\ref{Resolution}, extreme temperature values, in excess of $10000$\,K can be reached since most of the loaded energy is deposited within a shallow layer over only 1\,ms implying extremely high power density values. The MEMENTO code employs state-of-the-art analytical fits for all relevant W thermophysical properties from the room temperature up to very high temperatures (above the melting point and close to the normal boiling point)\,\cite{Tolias2017_1}, but above $\sim6000\,$K, all the fits have to be extrapolated. At extreme temperature values, the thermal conductivity falls sharply towards zero and eventually switches to unphysical negative values beyond $11500\,$K. This implies that no simulation can be carried out beyond $11500\,$K and that results with extreme temperature values must be viewed with caution. Further discussion may be found in Sec.\ref{sec:assessment}. 

General trends can be understood by analyzing the temporal evolution of the in-depth temperature profiles and of the interface temperature. Such results for some selected scenarios have been already presented in Fig.17 of Ref.\cite{Pitts_2025}. The crucial role of the vaporization cooling flux has been pointed out long ago\,\cite{Dabby_1972} and in more recent works\,\cite{Pitts_2025, Ratynskaia_2025b}. To be more specific, in spite of the monotonic energy deposition profile, the cells on the free surface effectively receive less energy compared to the bulk due to vaporization, resulting in a non-monotonic temperature profile and a maximum residing just underneath the surface. The latter is highlighted by the plots of the energy deposition and the resulting temperature shown in Fig.\ref{Resolution}. It is worth recalling that the actual energy deposition profile does feature a maximum, but for shallow angles, this is located in a sub-micron region underneath the surface and is not being resolved. As also pointed out earlier, the non-monotonicity of the temperature triggers PFC explosions due to the build-up of internal stresses underneath the surface. 

Although intense vaporization takes place during the loading time when the surface temperature is high, its effect can also be appreciated on the long timescale temperature response, namely in the maximum values of the coolant interface temperature. This is due to the vaporization losses, which can amount to a large fraction of the loaded energy, see Sec.\ref{sec:results_DINA}.  

\begin{figure*}
    \addtolength{\tabcolsep}{-10pt} 
    \begin{tabular}{cc}
        \subfloat{%
          \begin{overpic}[width=0.27\linewidth]{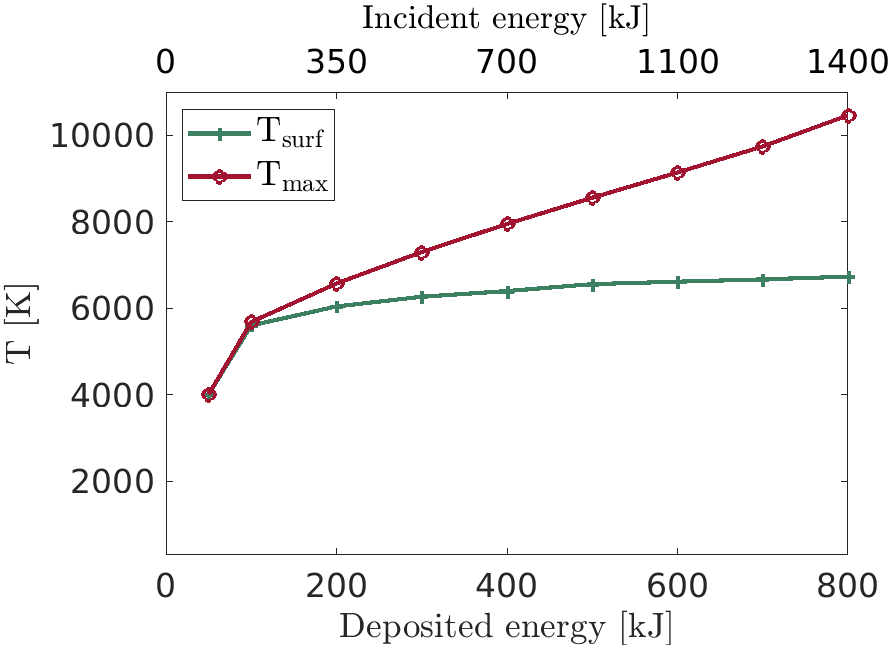}
            \put(83,27){{\textbf{\big(a)}}}
          \end{overpic}
        }
        \subfloat{%
          \begin{overpic}[width=0.245\linewidth]{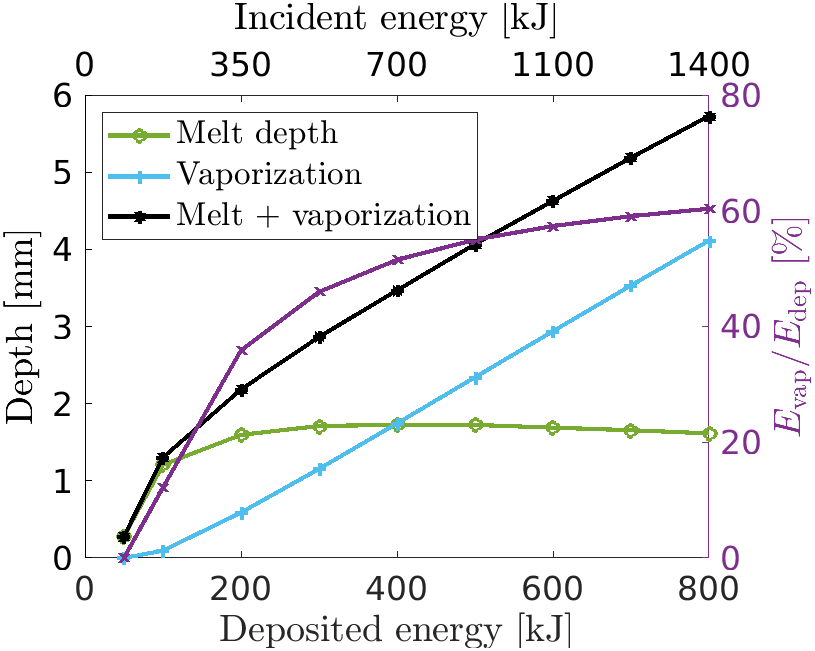}
           \put(73,31){{\textbf{\big(b)}}}
          \end{overpic}
        }
        \subfloat{%
          \begin{overpic}[width=0.25\linewidth]{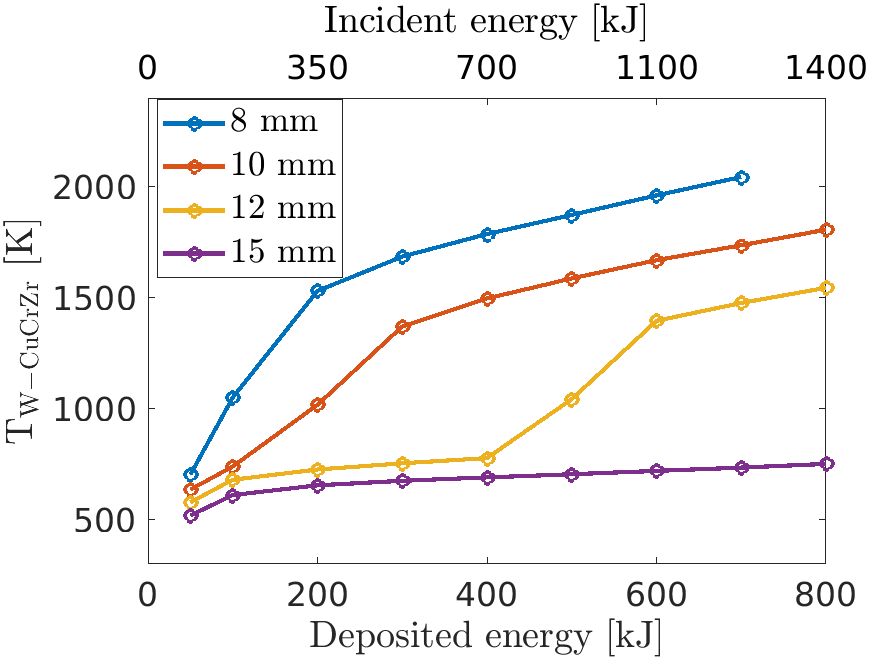}
          \put(83,27){{\textbf{\big(c)}}}
          \end{overpic}
        }
        \subfloat{%
          \begin{overpic}[width=0.25\linewidth]{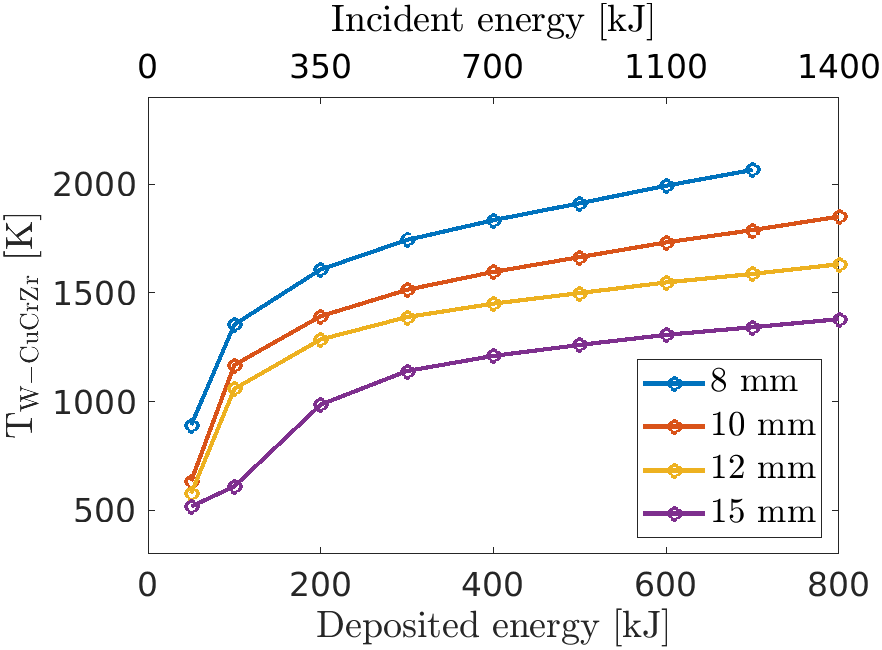}
          \put(30,15){{\textbf{\big(d)}}}
          \end{overpic}
        }
    \end{tabular}
    \addtolength{\tabcolsep}{6pt} 
    \caption{Surface damage (a-b) and cooling system response (c-d) in the DINA scenario for 100\,ms loading. The temperature at the W-CuCrZr interface is shown for the CHF at 30\,MW/m$^2$ in (c) and 10\,MW/m$^2$ in (d).}
    \label{fig:damage_100ms}
\end{figure*}

\subsection{Surface damage independence on the W armour thickness}

Simulations for all scenarios considered here reveal that the surface damage is insensitive to the W tile thickness. This stems from the combination of two aspects; (i) as discussed above, for the grazing RE incidence (in the relevant RE energy range), a major fraction of the energy is deposited in the near-surface layer leading to high energy density values, (ii) the damage occurs during the loading time when the maximum temperatures and temperature gradients are reached. The aforementioned heat diffusion length (120\,$\mu$m during 1\,ms loading and $1.2\,$mm during 100\,ms loading) implies that a W thickness beyond a few such lengths cannot affect the thermal response during loading. Hence, results for the surface temperature, the maximum temperature, the melt layer thickness and the vaporized layer thickness are the same for the 8, 10, 12 and 15\,mm W armour thickness.

\section{Results: DINA scenarios}\label{sec:results_DINA}

In this section, all plots feature double horizontal axes, which measure the incident energies as well as the deposited energies onto the panel. The 43$\%$ difference corresponds to the backscattering and transmission of REs and cascade products to the ambient, as discussed in Sec.\ref{sec:secondary_products}.

In the case of 1\,ms loading time, the power densities are so high that 11500\,K is reached once the deposited energy exceeds 50\,kJ, hence only relatively low energies could be probed. In the case of 100\,ms loading time, the power density values are a factor of $100$ lower and simulations of the thermal response up to 800\,kJ were possible.

\subsection{Loading over 1 ms}

The surface damage characteristics for this case are summarized in Figs.\ref{fig:damage_1ms}(a-b), with the curves fully overlapping for all thickness values.  Fig.\ref{fig:damage_1ms}(a) reveals that, with the exception of extremely low energies where vaporization is of no importance, the maximum temperature is larger than the surface temperature, implying that it resides within the bulk (see also the in-depth profile in Fig.\ref{Resolution}). Furthermore, the surface temperature almost saturates as vaporization becomes an important cooling process for higher energies. The erosion depth increases with the energy much faster than the surface temperature mainly due to the exponential sensitivity of the vapor pressure to $T_{\mathrm{surf}}$ and also because higher $T_{\mathrm{surf}}$ is reached earlier in the loading stage. The purple curve in Fig.\ref{fig:damage_1ms}(b) also shows how much of the deposited energy has been expended in vaporization, with the highest value of 20$\%$ naturally corresponding to 50\,kJ.

In the long timescale thermal response, given the low energies probed for 1\,ms, the CHF was not reached. We thus present the cooling system response in two different ways that highlight the effect of the deposited energy and the effect of the tile thickness, see Figs.\ref{fig:damage_1ms}(c) and (d), respectively. The dependence of the interface temperature $T_{\mathrm{W-CuCrZr}}$ on the energy deposited does not reveal an obvious scaling. The effects at play are more pronounced for 100\,ms and will be discussed in the following subsection. Fig.\ref{fig:damage_1ms}(d) shows that increasing the thickness has diminishing returns - for 15\,mm the $T_{\mathrm{W-CuCrZr}}$ values are closer to each other than at 8\,mm. With increased thickness, the interface temperature is evaluated further away from the heated near-surface volume and the dependence on the distance is weaker than linear. In fact, for larger thickness the maximum value is reached later.

\subsection{Loading over 100 ms}

The surface damage characteristics for this case are summarized in Figs.\ref{fig:damage_100ms}(a-b), with the curves now nearly overlapping for all thickness values. There is a minuscule (inconsequential) difference at higher energies due to the significant material erosion which means that the cooling interface is located closer to the loaded volume, thus affecting the temperature gradients. The effect of the increased loading time is observed when comparing $T_{\mathrm{surf}}$ and $T_{\mathrm{max}}$ in the 50\,kJ case; with $\sim7000$\,K and $\sim9500$\,K reached during the 1\,ms loading while only 4000\,K is reached during the 100\,ms loading. Note that, beyond the first two energy points in the scan, $T_{\mathrm{surf}}$ nearly saturates while the vaporization losses grow rapidly for the same reasons as in the 1\,ms case. The longer loading time is also responsible for the significantly larger vaporized layer at similar surface temperatures: compare the $h_{\mathrm{vap}}$ of 0.1\,mm in the 1\,ms case with the $h_{\mathrm{vap}}$ of several mm for 100\,ms. In fact, the vaporization losses amount up to 60$\%$ at larger energies, see the purple curve in Fig.\ref{fig:damage_100ms}(b). Moreover, from $\sim$300\,kJ, the melt depth saturates and starts falling down as erosion is faster than the melt creation from heat diffusion. Clearly, the total damage, quantified as vaporization plus melt, see the black curve in Fig.\ref{fig:damage_100ms}(b), becomes much worse at higher energies. 

The cooling system response is shown for the CHF at 30\,MW/m$^2$ and 10\,MW/m$^2$ in Figs.\ref{fig:damage_100ms}(c) and (d), respectively. The CHF of 30\,MW/m$^2$ is attained for energies above 50\,kJ for thickness of 8\,mm, above 100\,kJ for 10\,mm and above 400\,kJ for 12\,mm.  The CHF of 10\,MW/m$^2$ is reached above 50\,kJ, with the exception of 15\,mm where it is attained after 100\,kJ. The missing point for 8\,mm is due to numerical issues\,\footnote{MEMENTO can operate when different materials exist only in the lowest (most coarse) AMReX level. In this case, a large amount of material was eroded so that the CuCrZr layer was remeshed to the highest level. This does not affect the results for the surface damage.}.

The dependence of $T_{\mathrm{W-CuCrZr}}$ on the loaded energy is rather intricate. As the deposited energy increases, the vaporization cooling becomes important, implying that the dependence should be sub-linear. However, when significant erosion takes place, the energy is deposited closer to the cooling interface. The moving boundary complicates the problem. Finally, we recall that the heat transfer coefficient has a strong dependence on the interface temperature.

\section{Results: JOREK scenarios}\label{sec:results_JOREK}

\subsection{Energy deposition}

The results for the most loaded region (central part of the wetted area of the JOREK simulations) in each of the D1-D4 scenarios are presented in the upper row of Fig.\ref{fig:JOREK_Geant4_results} as depth-integrated power density 2D maps. These permit the most loaded tile to be identified, demarcated by the red boundaries in the figures.
Further information is provided in Table \ref{tab:JOREKtab}, where, in addition to the energy deposited per most loaded tile, the maximum energy density values are reported. It can be observed that while the energies per tile are very similar in all $4$ scenarios, the energy densities in the cases D3-D4 are nearly 20 times larger than those for D1-D2. This is due to the large difference in RE energy distributions. For D3-D4, they are characterized by an average value which is more than a factor $10$ lower than the D1-D2 mean RE energy. Such low RE energies result in ultra shallow energy deposition and correspondingly very high energy density values. 

It is also worth pointing out that the impact angles in the D4 scenario were appreciably larger than in the D1-D3 scenarios (compare the plots in the lower row of Fig.\ref{fig:JOREK_Geant4_results}). However, with typical RE energies of $\leq 2$\,MeV, this does not make the in-depth energy density profile more relaxed, hence D4 still features very high energy density values. As a consequence, see Sec.\ref{sec:grazing}, extreme temperatures are promptly reached and no meaningful simulations were possible for the D3-D4 scenarios.

\subsection{Thermal response}

\begin{figure}
    \centering
    \includegraphics[width=0.9\linewidth]{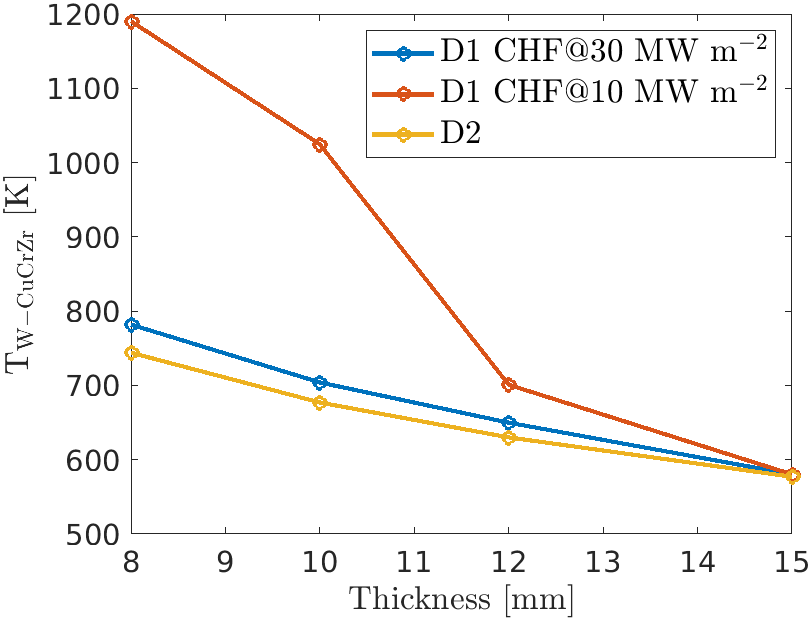}
    \caption{The temperature at the W-CuCrZr interface for different values of the W thickness for the JOREK scenarios D1 and D2. For D1, results are shown for two CHF values.}
    \label{fig:JOREK_thermal}
\end{figure}

Table \ref{tab:JOREKtab} summarizes the damage results for the JOREK scenarios D1-D2. The D1 case features a non-monotonic in-depth temperature profile due to the modest vaporization losses, while the D2 case is characterized by lower surface temperature and four times shorter loading time, hence vaporization cooling is not of importance and the temperature profile is monotonic. The interface temperatures are plotted in Fig.\ref{fig:JOREK_thermal} as a function of the tile thickness. For the D1 case, two curves are shown: for the CHF at 30 MW/m$^2$ and at 10 MW/m$^2$. 

In terms of the total energy loaded and the RE-PFC interaction time, the JOREK D1 scenario is similar to the DINA case of '50 kJ over 1 ms'. Translating to the most loaded tile leads to 10.5\,kJ in the D1 scenario (see Table \ref{tab:JOREKtab}) and to 6.7\,kJ in the DINA scenario. Concerning the surface damage, despite the larger deposited energy, the D1 case has milder PFC response than in the DINA scenario with lower in-depth and surface temperatures and less vaporization loss, albeit with somewhat larger melt thickness. However, the cooling system response is significantly worse, with the interface temperatures well above the maximum $T_{\mathrm{W-CuCrZr}}$ of $\sim 650$ K predicted for the DINA 50\,kJ case. The reason lies with the difference in the RE energy distributions; the larger population of low energy REs, below the mean value of 15 MeV, in the DINA scenario means that the energy density profile is quite steep. Fig.\ref{Resolution}(a) shows that the energy densities fall by factor of 20 within 0.5\,mm. The mono-energetic 26 MeV REs in the D1 case deposit energy deeper in the material, so that the profile is more relaxed and falls a mere factor of 2.5 over the first 0.5\,mm. The worse cooling system response is due to the energy being deposited closer to the cooling interface. 

\section{Secondary products} \label{sec:secondary_products}

A summary of all escaping products of the combined nuclear-electromagnetic shower is provided in Table \ref{tab:Losses}. The findings are somewhat scenario dependent and can be explained by the general physics arguments provided below.

The electron contribution to the energy loss can be substantial or minor. The strength of the contribution depends on the incident angle (with grazing angles drastically increasing the backscattering yield of primaries), the panel geometry (due to the possibility of transmission of primaries or deltas), the magnetic field strength \& topology (due to the possibility of prompt re-deposition in the course of the Larmor gyration) and naturally the incident energy (which is the upper bound of the exit energy distribution)\,\cite{Tolias2020,Komm2020,Reimer_2000}. It should be emphasized that grazing electrons are more likely to be immediately backscattered, which translates to exit energy distributions that are peaked at the incident energy\,\cite{Reimer_2000}. These are suprathermal electron populations that are continuously re-entering the plasma and which could therefore affect the RE dynamics.

The photon contribution to the energy loss is always substantial for MeV range REs in W. It should be emphasized that at incident energies roughly above $10.5\,$MeV in W, RE energy dissipation due to Bremsstrahlung overcomes the energy dissipation due to electron excitation and ionization\,\cite{Berger_1970,Berger_1992}. Thus, copious amounts of Bremsstrahlung photons are produced which are then scattered elastically (Rayleigh), scattered inelastically (Compton), converted (pair production) or absorbed (photoelectric effect or photo-nuclear reactions)\,\cite{PENELOPE_2018}. The photons generally have larger mean-free-paths than electrons but their exit energy distributions peak at low energies due to the efficiency of the inelastic channels at high energies\,\cite{Hubbell_1980}.

The positron contribution to the energy loss is negligible. Photon conversion to an electron-positron pair has a $\sim1.02\,$MeV threshold\,\cite{Motz_1969} and it becomes the dominant photon interaction in W for photon energies that exceed $\sim5\,$MeV\,\cite{Hubbell_1980}. However, positrons are second generation particles that are exclusively produced via the pair conversion of Bremsstrahlung photons in the field of the nuclei but also in the field of the electrons\,\cite{Seltzer_1986}. More importantly, they lose their energy efficiently in electronic stopping, nuclear stopping and electron-positron annihilation\,\cite{Berger_1983}, which prevents their escape to the ambient.

The neutron contribution to the energy loss is also generally negligible. Neutrons are mainly second generation particles, being primarily produced via the absorption of bremsstrahlung photons by nuclei (photo-nuclear reactions)\,\cite{Dietrich_1988}. They can also be first generation particles through electron-nuclear reactions or third generation particles through positron-nuclear reactions, but these nuclear paths are far less effective since the nucleus is excited by virtual photons\,\cite{Sari_2023}. Photo-nuclear W reactions have a kinematic threshold of $\sim8\,$MeV and a maximum cross-section around $\sim15\,$MeV\,\cite{HandbookIAEA_2000,Kawano2020}. Thus, photo-neutrons are generally produced in copious amounts and also have high escape probabilities. Nevertheless, the photo-neutrons have typical energies within $0.5-1.5\,$MeV\,\cite{Sari_2023}, which is largely responsible for the small associated energy loss. The statistics for both produced and escaping neutrons is provided in Table \ref{tab:Losses}, revealing that, independent of the simulated scenario, around 80$\%$ of neutrons produced escape the tile volume.

\begin{table*}[!h]
\centering
\renewcommand{\arraystretch}{1.2}
\begin{tabular}{m{6.2cm} m{1.5cm} m{1.5cm} m{1.5cm} m{1.5cm} m{2.0cm} m{2.0cm} }
\hline

\textbf{Scenario}  & \textbf{DINA}  & \textbf{D1} &\textbf{D2} & \textbf{D3} & \textbf{D4} \\
\hline

\textbf{Total energy loss, \%} & 42.73 & 22.52 & 18.83& 16.02 & 16.28 \\
 
\textbf{Contribution of electrons, \%} & 22.43 & 4.74 & 2.39 & 4.08 & 3.57 \\
 
\textbf{Contribution of positrons, \%} & 0.090 & 0.019 & 0.008 & 0.007 & 0.003 \\

\textbf{Contribution of photons, \%} &  20.20 &  17.74 & 16.42 &  11.93 &  12.71 \\

\textbf{Contribution of neutrons, \%} &  0.007 & 0.013 & 0.012 &  0.0026  &  0.0025 \\

\textbf{Neutron lost per $N$ electrons} &  $\sim700$ & $\sim230$ & $\sim270$ & $\sim45700$ & $\sim 52600$ \\

\textbf{Neutron produced per $N$ electrons} &  $\sim610$ & $\sim200$ & $\sim220$ & $\sim35700$ & $\sim 41000$ \\

\hline
\end{tabular}
\caption{Summary of secondary products emanating from the FW panels in the DINA and JOREK D1-D4 scenarios. The total energy loss, in \% of the energy loaded, is due to losses in electron backscattering, positron backscattering, escaping photons and escaping neutrons. Statistics for escaping and produced neutrons are also indicated.}
\label{tab:Losses}
\end{table*}

\section{Critical assessment of the results}\label{sec:assessment}

In spite of the rigorous modeling of the RE energy deposition and the PFC thermal response, there are various underlying assumptions that need to be taken into account when assessing the results.

{\textit{Specifications of the RE impact parameters.} The present ITER W FW thermal response results, similar to the recent DIII-D graphite dome thermomechanical response results\,\cite{Ratynskaia_2025}, reveal the acute sensitivity of the material damage to the details of the RE wall loading characteristics. We followed two approaches for the loading input, both having in common that the RE beam termination happens at $q_{95} \sim 2$ with $I_{RE}\sim 9$\,MA and that RE energy distributions are stipulated. In the DINA scenarios, the wetted area was prescribed by $\Delta_{RE}$, representing the most conservative assumption in terms of the localization of the RE energy deposition (given by the minimum width fixed by the Larmor radius). The loading times were also prescribed and the incident energies were scanned. Meanwhile, in the JOREK scenarios, RE loading is given by a strong single MHD driven termination, featuring much broader energy deposition, similar to those observed in benign termination experiments\,\cite{Hollmann2023NF}, and self-consistent RE-PFC interaction time and area. We also point out that the interaction of the REs with the backscattered or transmitted primary or secondary electrons emanating from the PFC (see Sec.\ref{sec:secondary_products}) can potentially affect the RE seeding, RE avalanches or RE termination processes. Modeling of such phenomena requires an iterative work-flow and will be the subject of future effort\,\cite{Ratynskaia_2025b}. 

\textit{Evolution of the free surface.} The RE impacts in the DINA and JOREK scenarios concern a pristine surface. However, the free surface is rapidly evolving, due to the (considered) vaporization losses, as well as the (neglected) melt motion and material explosion. As a result, incident REs may no longer strike the deformed surface or may strike at different angles, modifying the energy deposition profiles. The clear way to address this issue is to dynamically couple the MC and thermomechanical simulations, which is computationally costly. The thermo-mechanical response must be modeled first, as outlined below. 

Here, since only the thermal response is considered, we attempted to compensate for the rapid loss of vaporized material by shifting the energy density maps so that they follow the free surface evolution. To quantify the effect, a comparison with simulations without map shifting has been carried out for the case of 12\,mm thickness and 800\,kJ deposited energy in the DINA scenario. The results, presented in Table \ref{tab:Tabshift}, show that the mismatch between the Geant4 map for the pristine domain and the eroded MEMENTO domain results in the deposition of only a fraction of the energy, namely 42\,kJ out of 114\,kJ. As a consequence, in the simulations without map shifting, a layer of $<1\,$mm is vaporized compared with $>4\,$mm in the runs in which the map follows the free surface. It is worth pointing out that the slower erosion due to vaporization in the mismatched domain simulations allowed for a deeper melt layer compared to the shifted map simulations. Finally, it is observed that the interface temperatures are nearly identical. This is due to the fact that in the shifted map simulations, vaporization cooling is so efficient that $<50\%$ of the deposited energy was absorbed (44\,kJ out of 114\,kJ), making it comparable with the energy absorbed in the mismatched domain simulations (25 kJ). Moreover, in the former case, 4\,mm out of the nominal 12\,mm is lost, while in the latter case 11\,mm survives after the end of loading. 

\begin{table}[h!]
\centering
\begin{tabular}{lll}
\hline
 & Without shifting & With shifting \\
\hline
$T^{\text{surf}}_{\text{max}}/T_{\text{max}}$ [K]& 7172/6273 & 10470/6731 \\
$h_{\text{vap}}$ [mm] & 0.92 & 4.1 \\
$h_{\text{melt}}$ [mm] & 2.4 & 1.6 \\
$T_{\text{W--CuCrZr}}$ [K]& 1500 & 1544 \\
$E_{\text{dep}}/E_{\text{abs}}$ [kJ]& 42/25 & 114/44 \\
\hline
\end{tabular}
\caption{Comparison of key parameters with and without shifting of the power density map to compensate for the evaporated material losses.}
\label{tab:Tabshift}
\end{table}

\textit{Material explosion.} The assessment of the PFC damage and the cooling system response is exclusively based on the thermal response, neglecting mechanical and hydrodynamic aspects. Thus, the fact that the explosion would release a fraction of the deposited energy cannot be taken into account. Such an effect, whilst potentially releasing a substantial quantity of high velocity debris, would be beneficial with regard to the temperature at the cooling channel interface, in the same way as that afforded by vaporization losses. On the other hand, if an explosion does not take place, or if the melt layer is only partially disintegrated, the molten pool will not remain static as treated here. For example, it cannot be excluded that surface tension will force the W melt to drip over the corners and flow into the gaps, so that the hot melt might reach the CuCrZr layer or arrive in its vicinity, thus increasing the interface temperature. 

\textit{Thermo-physical properties.} Unavoidable uncertainties in the W thermo-physical properties at the extreme temperature range above the normal boiling point lead to hard-to-quantify inaccuracies, but do not undermine the overall conclusions with respect to the predicted damage. These results still serve the important purpose of being indicative of the dramatic damage to the upper layers of the material either via colossal vaporization when the temperature peak lies at the surface (owing to the exponential scaling, the vaporization rate is 0.5\,mm/ms at 8000\,K and 22\,mm/ms at 11000\,K) or via material explosion when the temperature peak resides underneath the surface.

\section{Conclusions and Outlook}\label{sec:conclusions}

\begin{figure}
    \centering
    \includegraphics[width=1.0\linewidth]{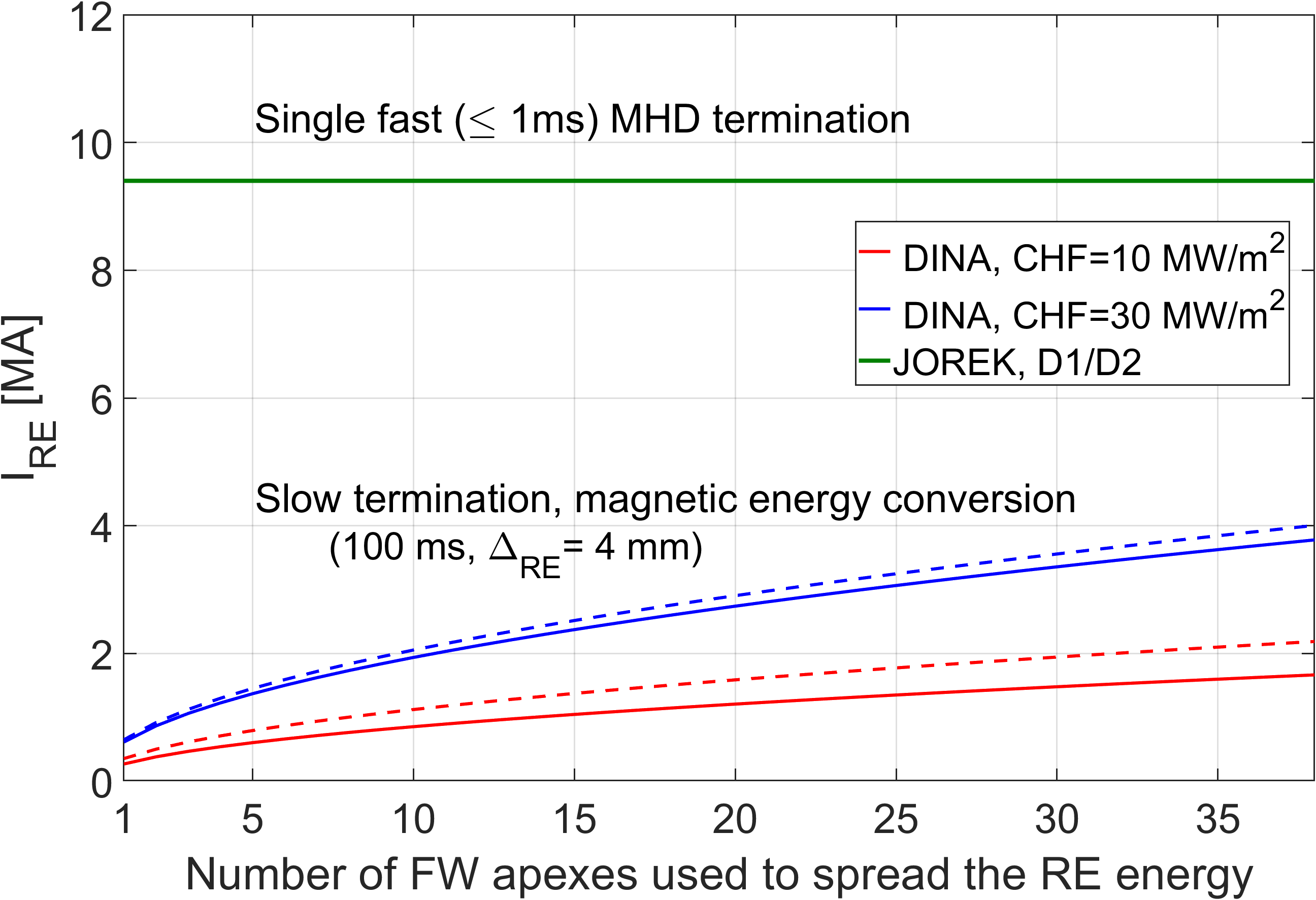}
    \caption{Damage thresholds, expressed as RE current values at which $T_{\mathrm{W-CuCrZr}}=1000$ K (full lines) or $T_{\mathrm{W-CuCrZr}}=1200$ K (dashed lines) is reached for a 12 mm W thickness. The calculations of curves corresponding to DINA scenarios are based on: (i) total incident energy scales as $E_{inc} \sim {I^2_{RE}}$ as per magnetic energy conversion, (ii) for $I_{RE}=9.4$ MA, $E_{inc}$  = 200 MJ, as obtained from JOREK simulations ~\cite{Bandaru_2025} and (iii) the total RE energy is evenly spread among the FW apexes. The green curve corresponds to the RE current that would produce $T_{\mathrm{W-CuCrZr}}=1000$ K in the case of the single fast MHD terminations simulated by JOREK. Based on Fig. \ref{fig:JOREK_thermal} for CHF$=10$ MW/m$^2$, the D1 and D2 cases are considered to be already marginal at 12 mm.}
    \label{fig:I_RE_plot}
\end{figure}

A three-stage, one-way coupled workflow, involving DINA and JOREK simulations of RE loading, Geant4 Monte Carlo simulations of RE volumetric energy deposition, MEMENTO simulations of heat transfer, has been employed for the predictive modeling of the ITER tungsten first wall thermal response to RE beam incidence. In the scenarios considered, the combination of RE energies and impact angles leads to ultra-shallow energy deposition, resulting in extreme temperatures and in vaporization induced, non-monotonic in-depth temperature response. Such temperature profiles, with the maximum residing underneath the surface, cause internal stress build-up and possibly material failure, fragmentation and fast debris expulsion. These phenomena have been observed in multiple machines due to accidental RE impact events\,\cite{Ratynskaia_2025b} and recently in the first controlled RE impact experiments with graphite in DIII-D\,\cite{Hollmann2025} and with W in WEST\,\cite{ITPA_Corre}.

From the modeling point of view, the ITER relevant RE impact regime proved challenging for the MC simulations given the high resolution requirement combined with the computationally demanding runs necessary to cover relatively large areas of the FW panels that were necessary to identify the most highly loaded sub-region (corresponding to the dimension of single, small castellations in the case of enhanced heat flux panels) in the presence of non-uniform loading due to the magnetic field. It also proved challenging for the heat transfer simulations due to the rapid loss of material by vaporization and the combination of high spatial resolution - short time simulations for the damage with long time simulations to obtain the temperature at the cooling interface. 

The thermal analysis of the short time response (over loading times from 0.25 to 100\,ms when the maximum temperature and temperature gradients are reached), clearly demonstrated that the surface damage is independent of the W tile thickness. This is caused by the shallow energy deposition and the short loading time, which imply that the characteristic diffusion lengths are sufficiently small compared to the W thickness (8 to 15\,mm). For high deposited energies, very intense vaporization occurs, leading to several mm of eroded material. On the other hand, such a rapid propagation of the vaporization front prevents deep melt layer formation, which stagnates below 2 mm in thickness. Since the damage extends to several mm, increased tile thickness can help to withstand more than a single event in a realistic scenario. Given the extremely severe penalty involved in opening the ITER torus and exchanging a complex FW panel, together with the need for some margin for incomplete disruption mitigation success, there is considerable potential gain in increasing thickness. Of course, the benefits of increased thickness for RE resilience always need to be offset against the need for adequate stationary heat exhaust under normal plasma operation conditions and the difficulty and cost of manufacturing extremely thick armour on the scale needed for the ITER FW ($\sim600\,$m$^2$ of surface area).

The thermal analysis of the long time response ($\sim1$\,sec) involves active cooling, which is incapacitated once the CHF values are reached. Our coolant interface temperature results reveal that the increased thickness provides more margin for RE-impact events before the W-CuCrZr interface is compromised. Moreover, while the vaporization is detrimental for the surface,  it initially assists in lowering the coolant interface temperature by removing a large fraction (more than 50\%) of the deposited energy. However, this benefit is lost when the erosion depth reaches extreme values due to the proximity of the energy deposition volume to the coolant. It is worth repeating that debris expulsion, not modeled here, has the same effect due to the removal of the thermal energy of the exploded volume.

Concerning the modeling of RE loading, two approaches have been pursued in this work. One of the motivations was the quantification of the effect of more relaxed impact angles of the JOREK simulations compared to the postulated zero pitch angle $5^\circ$ impacts in the DINA scenario. However, this effect was not possible to isolate due to the fact that the DINA and JOREK scenarios featured very different RE energy distributions. For cases of similar deposited energy on the most loaded tile, the JOREK scenario with 26\,MeV REs shows milder surface damage due to the less steep energy deposition profile, but higher cooling interface temperature due to the energy deposition closer to the coolant. Meanwhile, the low RE energy JOREK scenarios D3-D4 are associated with very high energy density values and hence extreme temperatures due to ultra-localized energy deposition in the near-surface layer, in spite of the similar deposited energies.  

Cooling interface limits can be cast in terms of the RE current, utilizing results presented in Figs.\ref{fig:damage_100ms} and \ref{fig:JOREK_thermal}. To be more specific, Fig.\ref{fig:I_RE_plot} shows at which RE currents the interface temperatures reach a critical value between $T_{\mathrm{crit}}=1000$\,K and $1200$\,K, given different assumptions on the number of affected panel apexes. This critical interface temperature is not a well defined number, since even empirical tests of the consequences of such intense transients as those delivered by RE impacts do not exist. Specific issues such as the softening of the substrate at higher temperature leading to potential delamination of the W armour are possible causes for concern. However, given the  ${I_{RE}}\sim\sqrt{E_{inc}}$ scaling, the variation between two choices of the critical temperature is not significant, as seen from the comparison of solid and dashed lines in Fig.\ref{fig:I_RE_plot}. The results are highly sensitive to the termination type and the loading assumptions. For single, fast MHD terminations, RE currents up to 9\,MA may not compromise the integrity of cooling channels. In contrast, slow terminations characterized by Larmor radius scale deposition areas and magnetic energy conversion, yield much lower thresholds: 1.5-3.7\,MA when the RE energy is evenly distributed over a full toroidal row of FW panels, which equates to a total of 36 apexes in the current ITER FW design for the rows expected to be most concerned by the RE impacts, and only 0.3-0.6\,MA when all the energy is concentrated on a single apex.

Finally, concerning future work, there is ongoing effort to extend the thermal code-chain of the workflow into a full thermomechanical code-chain that includes the W hydrodynamic and viscoplastic responses\,\cite{Ratynskaia_2025b}. It is emphasized that the inclusion of an equation of state over an extended range of pressures and temperatures would also solve the problem arising here with the extrapolated thermophysical properties at extreme temperatures in excess of 10000\,K.  Furthermore, it is pointed out that the recent controlled WEST experiment on RE-induced damage of a W tile and a similar planned upcoming controlled experiment in ASDEX Upgrade will provide valuable empirical data for the validation of the complete hydrodynamic / thermomechanical model.

\section*{Acknowledgments}

\noindent This work has been carried out under ITER Organization Service Contract IO/24/CT/4300003101. ITER is the Nuclear Facility INB No.\,174. This publication is provided for scientific purposes only and its contents should not be considered as commitments from the ITER Organization as a nuclear operator in the frame of the licensing process. The views and the opinions expressed herein do not necessarily reflect those of the ITER Organization. The Geant4 and MEMENTO simulations were enabled by resources provided by the National Academic Infrastructure for Supercomputing in Sweden (NAISS) at the NSC (Link\"oping University) partially funded by the Swedish Research Council through grant agreement No\,2022-06725.\\

\bibliography{biblio_new}
\end{document}